\documentclass[12pt,a4paper]{article}
\usepackage[dvipsnames]{xcolor}
\usepackage[export]{adjustbox}
\usepackage{xcolor}
\usepackage{color,soul}
\usepackage{amsmath,amsfonts,amssymb}
\usepackage{graphicx,rotate,multicol,wrapfig}
\usepackage{color}
\usepackage{float}
\usepackage{epsfig,epsf}
\usepackage{bm}
\usepackage{dcolumn,subfig}
\usepackage{array}
\usepackage{relsize}
\usepackage[bookmarks, breaklinks, colorlinks,urlcolor=blue, citecolor=red, 
linkcolor=blue]{hyperref}
\usepackage[font=small,labelfont=bf]{caption} 

\usepackage{caption}
\usepackage{multirow}
\usepackage{graphics}
\usepackage{hyperref}
\usepackage{graphicx} 
\usepackage{amsmath}
\usepackage{amsfonts}
\usepackage{amssymb}
\usepackage{bigints} 
\usepackage{relsize} 
\usepackage[top=1in, bottom=2cm, left=1.5cm, right=1.5cm]{geometry}
\usepackage{authblk}
\usepackage{changepage} 
\usepackage{array, makecell} %
\usepackage{afterpage}
\usepackage{placeins}
\usepackage{wrapfig}
\usepackage{soul}
\usepackage{cancel}
\usepackage{url}
\usepackage{braket}
\usepackage{hyperref}
\usepackage{amsfonts}
\usepackage{amssymb}
\usepackage{bigints} 

\usepackage{authblk}
\usepackage{cite}
\usepackage{multirow}
\usepackage{changepage} 
\usepackage{array, makecell} %
\usepackage{afterpage}
\usepackage{placeins}
\usepackage{titlesec}
\usepackage{caption}
\usepackage{wrapfig}
\usepackage{soul}
\usepackage{cancel}
\usepackage{url}
\usepackage{acronym}

\def\beq{\begin{equation}}
\def\eeq{\end{equation}}
\def\bea{\begin{eqnarray}}
\def\eea{\end{eqnarray}}
\def\nn {\nonumber}

\DeclareUnicodeCharacter{2212}{\ensuremath{-}}
\usepackage[utf8x]{inputenc}

\def\Babar{{\mbox{\slshape B\kern-0.1em{\smaller A}\kern-0.1em B\kern-0.1em{\smaller A\kern-0.2em R}}}}
\begin{document}

\renewcommand*{\thefootnote}{\fnsymbol{footnote}}

\begin{center}
	{\Large\bf{ Impact of Vector like quarks on $(g-2)_{\mu}$ with X-II-2HDM scenario and its phenomenological implications.}}\\[5mm]
Md.Raju$^{a,}$\footnote{mdrajuphys18@klyuniv.ac.in}, 
Abhi Mukherjee$^{b,}$\footnote{abhiphys18@klyuniv.ac.in}, and
Jyoti Prasad Saha$^{a,}$\footnote{jyotiprasadsaha@gmail.com}\\[3mm] {\small\em $^a$Department of Physics, University of Kalyani, Nadia 741235, India}\\ {\small\em
$^b$Gokhale Memorial Girls' College, 1/1 Harish Mukherjee Road, Kolkata 700 020, India}	
\end{center}


\begin{abstract}
The recent observation of the muon $(g-2)_{\mu}$ anomaly continues to challenge the explanations provided by the Standard Model. However, this anomaly can potentially find reconciliation within the framework of two-Higgs doublet models, provided that the pseudoscalar mass remains low. The introduction of additional fermionic components, such as a generation of vector-like quarks, not only broadens the acceptable parameter range for elucidating the anomaly but also presents an opportunity to circumvent conflicts with constraints from B-decays and heavy Higgs searches. We demonstrate the efficacy of fitting the anomaly in the muon magnetic moment within these models, assuming that vector-like quarks do not undergo mixing with Standard Model quarks.  With interactions following a type-X pattern for standard model quarks and a type-II pattern for vector-like quarks, results in models designated as type-XII2HDMVLQ. Additionally, we have explored double Higgs production within this model and observed when both the heavy Higgs and VLQ contribute the double Higgs production cross section significantly enhanced.

\end{abstract}

\setcounter{footnote}{0}
\renewcommand*{\thefootnote}{\arabic{footnote}}
\section{Introduction}
The Standard Model (SM) offers a remarkable understanding of nature that endures harsh scrutiny at both the present energy and accuracy frontiers. Stringent bounds up to several TeV are imposed for several new particles due to the lack of any direct signal for them in the present day collider. The extraordinary agreement between the predictions made by the SM and the experimental data obtained so far from the LHC has shown that the SM is the most appropriate effective theory for electroweak (EW) symmetry breakdown. Although there are some experimental data that can not be explained by the SM. One notable example is the anomalous magnetic moment of the muon ($(g-2)_{\mu}$). The measured value of the magnetic moment of the muon deviates from the SM prediction by more than 3$\sigma$.  The  $(g-2)$ Collaboration of Fermilab  recently published a new result from Run 1 experiment measuring the anomalous magnetic moment of the muon  \cite{Muong-2:2021ojo,Muong-2:2021vma,Muong-2:2021ovs,Muong-2:2021xzz}. Before this result the discrepancy between the experimental measurement of the $(g-2)_{\mu}$ \cite{Muong-2:2006rrc} and the SM $(g-2)_{\mu}$prediction\cite{Aoyama:2020ynm} was at a standard deviation of 2.7$\sigma$. A precise analysis of the new combined experimental data reveals \cite{Muong-2:2021ojo} 
\begin{equation}
\Delta a_{\mu} = a_{\mu}^{exp} - a_{\mu}^{SM} =(249   \pm 48) \times 10^{-11} \hspace{2em} (3.7 \sigma)
\end{equation}
where we define $a_{\mu} = \Big(\frac{g-2}{2}\Big)_{\mu}$
while the new combined result is \cite{Muong-2:2021ojo}   observed value of $a_{\mu}$ deviates from the SM predicted value \cite{Aoyama:2020ynm} at 4.2$\sigma$ confidence level.

Numerous papers have proposed diverse explanations for the $(g-2)_{\mu}$ anomaly, encompassing a wide array of new physics models. Examples include supersymmetric models, left-right symmetric models, scotogenic models, 331 models, $L_{\mu} - L_{\tau}$ models, seesaw models, and the Zee-Babu model \cite{Joglekar:2013zya,Kyae:2013hda,PhysRevLett.44.912,Ma:2006km,Long:1995ctv,Heeck:2010pg,Schechter:1980gr,Zee:1985rj,Babu:1988ki,Lindner:2016bgg,Dermisek:2013gta,Falkowski:2013jya,Branco:2011iw,Bhattacharyya:2015nca}.

However, in the context of this paper, we specifically concentrate on Two-Higgs Doublet Models (2HDMs)\cite{Branco:2011iw,Bhattacharyya:2015nca}. Our emphasis lies in discussing the implications arising from the latest findings by the Muon $(g-2)$ Collaboration. We aim to shed light on the distinctive contributions and implications that this particular framework offers in the understanding of the $(g-2)_{\mu}$ anomaly, thereby contributing to the ongoing discourse on new physics explanations.

In the realm of 2HDM, it becomes feasible to eliminate tree-level flavor-changing neutral currents mediated by scalar particles through the imposition of a discrete $Z_2$ symmetry onto the model. Notably, there exist four distinct versions of 2HDMs: type-I, type-II, type-X, and type-Y (flipped) \cite{PhysRevD.15.1958}. However, among these variants, only type-II and type-X demonstrate effectiveness in elucidating the $(g-2)_{\mu}$ anomaly. Specifically, the type-II and type-X configurations feature an augmented coupling of leptons with the new heavy scalar of the 2HDM, offering a viable resolution to the muon anomaly. This resolution encompasses the incorporation of both the customary one-loop and two-loop contributions from Barr Zee-type diagrams \cite{Wang:2018hnw, Broggio:2014mna}.

The type-II model faces significant constraints from flavor physics and direct searches for additional Higgs bosons due to the proportional relationship between the coupling of the new heavy scalar with both charged leptons and down-type quarks and the parameter $\tan\beta$, where $\tan\beta$ represents the ratio of the two vacuum expectation values. In both type-II and type-X models, the coupling of the muon to the heavy Higgs bosons experiences an enhancement by a factor of $\tan\beta$. However, in type-II models, the quark couplings with $Q=-\frac{1}{3}$ also undergo the same enhancement, potentially giving rise to issues in the context of radiative B-meson decays. Notably, the challenge of explaining the $(g-2)_{\mu}$ anomaly without adversely affecting B decays serves as a key motivation for investigating the type-X model.

In the type-II model, achieving the $(g-2)_{\mu}$ anomaly necessitates a high value of $\tan\beta$ and a light pseudoscalar mass, a combination that, according to B-physics observables, is not permissible \cite{Chun:2015xfx}. Conversely, the flavor restrictions in the type-X 2HDM are comparatively milder than those in the type-II 2HDM, owing to enhanced lepton couplings and suppressed quark couplings \cite{Cao:2009as}. This characteristic allows the type-X model to exclusively align with explaining the existing muon anomaly while evading flavor constraints, without resorting to any fermionic extensions. However, a notable drawback of the type-X model emerges, as satisfying low-energy data requirements entails the necessity of an exceedingly light pseudoscalar Higgs boson and a substantial $\tan\beta$ \cite{Wang:2018hnw, Broggio:2014mna} \cite{Chun:2015xfx, Wang:2014sda, Abe:2015oca, Chun:2015hsa} conditions that are also incompatible with B-physics observables.

The parameter space of the type-X 2HDM model is significantly restricted by experimental measurements of $H \rightarrow \tau \tau$ decay, leading to the exclusion of the parameter region capable of explaining the discrepancy in $(g-2)_{\mu}$ at the $1\sigma$ level due to constraints from heavy Higgs searches \cite{Abe:2015oca}. To address these limitations, we propose an enhancement by introducing a single generation of vector-like quarks (VLQs) into the framework of the 2HDM. The model is called type X-II-2HDM+VLQ where the SM fermion coupled in type-X way and the VLQ as type-II 2HDM like.  Our analysis demonstrates that this modification effectively mitigates the aforementioned issues associated with the individual models, all the while offering a minimalistic approach to explaining the $(g-2)_{\mu}$.

 Notably, in this model, there is no   mixing between these additional VLQs and the SM quarks. But the charged and colord VLQ can effect the loop induced process like $h \rightarrow \gamma \gamma$ and $h \rightarrow g g $ which is experimentally well measured. So , we have demonstrate that the decay of the SM Higgs to $\gamma \gamma$ remains well within the bounds of experimental uncertainty, courtesy of contributions from VLQ loops and the loop involving the charged Higgs scalar. Importantly, the coupling of VLQs with non-SM Higgs bosons introduces additional Barr Zee contributions, augmenting the value of $(g-2)_{\mu}$ even in the presence of a heavier pseudoscalar Higgs boson. This configuration can be perceived as a minimalistic framework wherein VLQs effectively address the $(g-2)_{\mu}$ anomaly while adhering to theoretical and experimental constraints. As there is no direct decay of VLQs in SM fermions  so its direct siggnature cant not be probed. But the as we have said that the VLQs can arise in loop induced process. The double higgs propduction is one of the improtant place where such heavy fermion can effect. So any enhancement of double higgs production can be the possible hint of such fermion. In this work we have shown the enhacemant due the VLQs in double higgs production.

This paper is structured as follows: In Section \ref{model}, we explore the extended Type-X-II 2HDM+VLQ model, which is intricately connected with Vector-Like Quarks (VLQs), and elucidate its impact on the $(g-2)_\mu$. Section \ref{constraints} provides a comprehensive compilation of both experimental and theoretical constraints utilized to rigorously confine our model. The constraints outlined here serve as the foundation for assessing the viability of our proposed model. Moving forward, in Section \ref{muon}, we meticulously examine the contributions from pertinent one-loop and two-loop Barr Zee diagrams to the muon anomalous magnetic moment within this specific scenario. Subsequently, in Section \ref{result}, we delve into a detailed discussion of the results and findings derived from our investigation. In Section \ref{dihiggs}, we have discussed the production cross section of the VLQs at the LHC and how it depends on various parameters. We have also presented the distributions of different kinematic variables, which are crucial for further collider studies. Finally, Section \ref{conclusion} serves as the concluding segment, summarizing and wrapping up all the key discoveries and insights obtained throughout our study.
\section{The Model} \label{model}
Let's delve into a Type-X-II 2HDM+VLQ model with a CP-conserving scalar potential, where two doublets are denoted as $\Phi_1$ and $\Phi_2$. The vacuum expectation values (VEVs) of these doublets are represented by $v_1$ and $v_2$ respectively, with $v=\sqrt{v_1^2 + v_2^2 }$ denoting the total electroweak VEV. This scalar spectrum consists of two CP-even neutral fields, denoted as $h$ and $H$, with $h$ being the lighter one; one CP-odd neutral field, $A$; and a pair of charged fields, $H^{\pm}$. The physical fields $h$ and $H$ are derived from the corresponding CP-even components of $\Phi_1$ and $\Phi_2$ through a rotation parameterized by the angle $\alpha$. Introducing a single generation of VLQs along with the type X-2HDM forms a key aspect of our model. This new inclusion encompasses a set of quarks sharing identical quantum properties with the standard quarks. To designate these VLQs, we employ the following notation:

\begin{center}
\hspace{0.5em} \textit{Higgs doublets}:\hspace{0.2cm} $\Phi_i = 
\begin{pmatrix}
h_i^+ &\\
\frac{v_i + \mathsf{h}_i + i \mathsf{a}_i}{\sqrt{2}} 
\end{pmatrix} $ ,\hspace{0.4cm}$(i=1,2)$ \\
\vspace{0.2cm}
 \hspace{-0.5em}\textit{VLQ doublets}:  \hspace{0.2cm}   $\mathcal{Q}_L = 
\begin{pmatrix}
T^{'}_L &\\
B^{'}_L
\end{pmatrix} ,\hspace{0.2cm}
\mathcal{Q}_R = 
\begin{pmatrix}
	\mathcal{T}^{'}_R &\\
	\mathcal{B}^{'}_R
\end{pmatrix}$

\vspace{0.5cm}
 \hspace{0.0em} \textit{VLQ singlets}:  \hspace{0.2cm}  $\mathcal{T}_L ,  \hspace{0.2cm}\mathcal{T}_R , \hspace{0.2cm}\mathcal{B}_L , \hspace{0.2cm}\mathcal{B}_R$
\end{center}

The comprehensive Yukawa  Lagrangian for   type \textsc{X-II 2HDM+VLQ}:
\begin{equation}
\begin{split}
\label{eq:lagrangian}
\mathcal{L}_{\mathsf{y}}= -y_u Q_{L} \tilde{\Phi}_2 u_R - y_d Q_{L} \Phi_2 d_R - y_{\ell} L_{L} \Phi_1 e_R \\ 
-M_\mathcal{Q} \bar{\mathcal{Q}} \mathcal{Q} -  M_\mathcal{T} \bar{\mathcal{T}} \mathcal{T} - M_\mathcal{B} \bar{\mathcal{B}} \mathcal{B}  - Y_\mathcal{B} \bar{\mathcal{Q}} \Phi_1 \mathcal{B} - Y_\mathcal{T} \bar{\mathcal{Q}} \tilde{\Phi}_{2} \mathcal{T}  +  \mathsf{H.c}
\end{split}
\end{equation}
where $\tilde{\Phi}= i \tau_2 \Phi^{*}$ and we assume that $Y_{\mathcal{T}}^R = Y_{\mathcal{T}}^L = Y_{\mathcal{T}} $ and $Y_{\mathcal{B}}^R = Y_{\mathcal{B}}^L = Y_{\mathcal{B}} $  for simplicity.

\begin{table}[h]
	\begin{center}
		\begin{tabular}{||c| c| c| c| c| c| c| c||} 
		\hline
      \hline
	
  Model & $\Phi_{1}$ & $\mathcal{Q}_L$  & $\mathcal{Q}_R$ & $\mathcal{T}_L$ & $\mathcal{T}_R$ &  $\mathcal{B}_L$  &$\mathcal{B}_R$  \\ 
  \hline
  VLQs &    $+$&   $+$&  $+$  &   $-$ &   $-$&   $+$&   $+$\\ 
  \hline
  \hline
	\end{tabular}
		\caption{\label{sm-parameters}\textit{The $Z_2$ parities of the SM fermions and VLQs}}
	\end{center}
 \label{tab:quantum}
\end{table}
The Dirac and Yukawa interaction term yields the VLQs mass matrices. The $(\mathcal{B}^{'} \mathcal{B})$ and $(\mathcal{T}^{'} \mathcal{T})$ basis can be used to write the mass matrices as

\begin{equation} \label{eq:massmatrix}
\begin{split}
M_{\mathcal{B}} &=
\begin{pmatrix}
M_{\mathcal{Q}} && \frac{1}{\sqrt{2}}Y_{\mathcal{B}}\cos\beta \\
 \frac{1}{\sqrt{2}}Y_{\mathcal{B}} v \cos\beta  && M_{\mathcal{B}}   
\end{pmatrix} \\
\vspace{0.4cm}
M_{\mathcal{T}} &=
\begin{pmatrix}
M_{\mathcal{Q}} && \frac{1}{\sqrt{2}}Y_{\mathcal{T}}\sin\beta \\
 \frac{1}{\sqrt{2}}Y_{\mathcal{T}} v \sin\beta  && M_{\mathcal{T}}   
\end{pmatrix}
\end{split}
\end{equation}

The matrix $\mathcal{R}(\theta_{\mathcal{F}})$ performs a rotation on the VLQs, transforming them from their weak eigenbasis to the mass basis, while also diagonalizing the mass matrix.

\begin{equation}\label{eq:rotation}
\mathcal{R}(\theta_{\mathcal{F}}) = 
\begin{pmatrix}
\mathsf{c}_{\theta_{\mathcal{F}}} && -\mathsf{s}_{\theta_{\mathcal{F}}} \\
 \mathsf{s}_{\theta_{\mathcal{F}}} && \mathsf{c}_{\theta_{\mathcal{F}}}   
\end{pmatrix}
\end{equation}

where $\mathsf{\cos}\theta_{\mathcal{F}}=\mathsf{c}_{\theta_{\mathcal{F}}}$ and  $\mathsf{\sin}\theta_{\mathcal{F}}=\mathsf{s}_{\theta_{\mathcal{F}}}$.

\begin{equation}\label{eq:massbasis}
\begin{pmatrix}
    \mathcal{B}_1 \\
    \mathcal{B}_2
\end{pmatrix}=
R(\theta_{\mathcal{B}}) \begin{pmatrix}
    \mathcal{B}^{'} \\
    \mathcal{B}
\end{pmatrix}, \hspace{0.5cm}\begin{pmatrix}
    \mathcal{T}_1 \\
    \mathcal{T}_2
\end{pmatrix}=
R(\theta_{\mathcal{T}}) \begin{pmatrix}
    \mathcal{T}^{'} \\
    \mathcal{T}
\end{pmatrix}
\end{equation}

We can treat $\mathcal{T}_1$ and $\mathcal{B}_1$ are $SU(2)$ doublet, and $\mathcal{T}_2$ and  $\mathcal{B}_2$  as singlet when the values of $\theta_{\mathcal{T}}$ and  $\theta_{\mathcal{B}}$ are much smaller than 1.

Simplified notation such as $\mathsf{c}_{\mathcal{T,B}}=\mathsf{c}_{\theta_{\mathcal{T,B}}}$ and   $\mathsf{s}_{\mathcal{T,B}}=\mathsf{s}_{\theta_{\mathcal{T,B}}}$ can be used to express the VLQs rotation angle.
\vspace{-5mm}
\begin{equation}\label{eq:angle}
\begin{split}
s_{2\mathcal{B}} &= \frac{\sqrt{2}Y_{\mathcal{B}} v}{M_{\mathcal{B}_2}-M_{\mathcal{B}_1}}c_{\beta}\\
s_{2\mathcal{T}} &= \frac{\sqrt{2}Y_{\mathcal{T}} v}{M_{\mathcal{T}_2}-M_{\mathcal{T}_1}}s_{\beta}
\end{split}
\end{equation}

Since every kind of 2HDM Yukawa structure has previously been provided with an explicit coupling factor in Chapter(2), we are only representing the VLQ portion below.

\begin{equation}\label{eq:masslag}
\begin{split}
\mathcal{L}_{y}^{\textsc{VLQ}} &=  - \sum_{\mathcal{F}=\mathcal{T},\mathcal{B}}\sum_{\mathsf{i,j}=1}^{2} \sum_{\mathsf{\phi}=\mathsf{h,H}} y^{\mathsf{\phi}}_{\mathcal{F}_\mathsf{i} \mathcal{F}_\mathsf{j}} \mathsf{\phi} \bar{\mathcal{F}_\mathsf{i}} \mathcal{F}_\mathsf{j} - \sum_{\mathcal{F}} \sum_{\mathsf{i,j}=1}^{2} \big[ -i y^A_{\mathcal{F}_\mathsf{i} \mathcal{F}_\mathsf{j}}A \bar{\mathcal{F}}_{\mathsf{i},R} \mathcal{F}_{\mathsf{j},L}  +\mathsf{H.c}  \big] \\&
- \sum_{\mathsf{i,j}=1}^{2} \big[ y^{\mathsf{H}^{\pm}}_{\mathcal{T}_\mathsf{i} \mathcal{B}_j} \mathsf{H}^+ \bar{\mathcal{T}_i} \mathcal{B}_j + \mathsf{H.c}     \big]
\end{split}
\end{equation}

The Tab. \eqref{tab:couplingfactor} contains a comprehensive presentation of the entire set of form factors associated with the Yukawa Lagrangian.
\begin{table}[!htb]
	\begin{center}
		\begin{tabular}{||c|c|c|c|| } 
		\hline
  \hline
	  $\eta^h_{\mathcal{T}}= \eta^H_{\mathcal{B}}$  & $\eta^H_{\mathcal{T}}=  -\eta^h_{\mathcal{B}}$ &  $\eta^A_{\mathcal{T}}$ & $\eta^A_{\mathcal{B}}$  \\ 
		\hline
	         $\cos\alpha$ &  $\sin\alpha$    &    $\cos\beta$ & $\sin\beta$   \\
		\hline
  \hline
   $y^{\mathsf{\phi}}_{\mathcal{F}_\mathsf{1} \mathcal{F}_\mathsf{1}} = -y^{\mathsf{\phi}}_{\mathcal{F}_\mathsf{2} \mathcal{F}_\mathsf{2}}$ &  $y^{\mathsf{\phi}}_{\mathcal{F}_\mathsf{1} \mathcal{F}_\mathsf{2}} = -y^{\mathsf{\phi}}_{\mathcal{F}_\mathsf{2} \mathcal{F}_\mathsf{1}}$    &    $y^{\mathsf{A}}_{\mathcal{F}_\mathsf{i} \mathcal{F}_\mathsf{i}} $ & $y^{\mathsf{A}}_{\mathcal{F}_\mathsf{1} \mathcal{F}_\mathsf{2}} = -y^{\mathsf{A}}_{\mathcal{F}_\mathsf{2} \mathcal{F}_\mathsf{1}}$   \\
		\hline
	        $-\frac{1}{\sqrt{2}} Y_{\mathcal{F}} \eta_{\mathcal{F}}^{\mathsf{\phi}} \mathsf{s}_{2 \mathcal{F}} $ &  $\frac{1}{\sqrt{2}} Y_{\mathcal{F}} \eta_{\mathcal{F}}^{\mathsf{\phi}} \mathsf{c}_{2 \mathcal{F}}$    &    $0$ & $\frac{1}{\sqrt{2}} Y_{\mathcal{F}} \eta_{\mathcal{F}}^{\mathsf{A}}$   \\
		\hline
        \hline
	    $y^{\mathsf{H}^+}_{\mathcal{T}_1 \mathcal{B}_1}$ &  $y^{\mathsf{H}^+}_{\mathcal{T}_1 \mathcal{B}_2}$    &    $y^{\mathsf{H}^+}_{\mathcal{T}_2 \mathcal{B}_1}$ & $y^{\mathsf{H}^+}_{\mathcal{T}_2 \mathcal{B}_2} $   \\
		\hline
	         $-Y_{\mathcal{T}} \eta^\mathsf{A}_{\mathcal{T}} \mathsf{c}_{\mathcal{B}} \mathsf{s}_{\mathcal{T}} - Y_{\mathcal{B}} \eta^\mathsf{A}_{\mathcal{B}} \mathsf{s}_{\mathcal{B}} \mathsf{c}_{\mathcal{T}} $ &  $-Y_{\mathcal{T}} \eta^\mathsf{A}_{\mathcal{T}} \mathsf{s}_{\mathcal{B}} \mathsf{s}_{\mathcal{T}} + Y_{\mathcal{B}} \eta^\mathsf{A}_{\mathcal{B}} \mathsf{c}_{\mathcal{B}} \mathsf{s}_{\mathcal{T}}$    &    $Y_{\mathcal{T}} \eta^\mathsf{A}_{\mathcal{T}} \mathsf{c}_{\mathcal{B}} \mathsf{c}_{\mathcal{T}} - Y_{\mathcal{B}} \eta^\mathsf{A}_{\mathcal{B}} \mathsf{s}_{\mathcal{B}} \mathsf{s}_{\mathcal{T}}$ & $Y_{\mathcal{T}} \eta^\mathsf{A}_{\mathcal{T}} \mathsf{s}_{\mathcal{B}} \mathsf{c}_{\mathcal{T}}  + Y_{\mathcal{B}} \eta^\mathsf{A}_{\mathcal{B}} \mathsf{c}_{\mathcal{B}} \mathsf{s}_{\mathcal{T}}$   \\
		\hline
  \hline
\end{tabular}
		\caption{\label{2hdm-yukawa}\textit{The interaction factor for VLQ in Eq.\eqref{eq:masslag}.}}\label{tab:couplingfactor}
	\end{center}
 
\end{table}

The key observation is that the lighter CP-even scalar, $h$, exhibits couplings with the SM gauge bosons and fermions that align closely with those predicted by the SM (within experimental uncertainties). This alignment significantly constrains the permissible parameter space \cite{Gunion:2002zf,Carena:2013ooa,Bhattacharyya:2015nca,Bhattacharyya:2014oka,Bhattacharyya:2013rya}.

\begin{equation}
\cos(\beta -\alpha) \approx 0
\end{equation}

On the other hand, to cancel the VLQs contributions completely from the $gg \rightarrow h$, $h \rightarrow \gamma \gamma$, and $h \rightarrow Z \gamma$, processes, one needs the correlation among the scaling of the Higgs couplings relative to the SM,

\begin{equation}
\frac{y_{VVh}}{y_{VVh}^{SM}}= \frac{y_{uuh}}{y_{uuh}^{SM}} = - \frac{y_{ddh}}{y_{ddh}^{SM}} = - \frac{y_{\ell \ell h}}{y_{\ell \ell h}^{SM}} = 1
\end{equation}

where $V$, $u$, $d$, and $\ell$ denote, generically, the weak gauge bosons, the up-type quarks, the down-type quarks, and the charged leptons, respectively. Several sources have considered the possibility of a scenario known as the ``wrong sign limit'' in the 2HDM-II model, where the relative coupling between up and down quarks reverses its sign \cite{Fontes:2014tga,Ferreira:2014dya,Biswas:2015zgk,Ferreira:2014naa}.  The ``wrong sign limit" also appears between up and down type VLQs in our model type X-II-2HDM+VLQs \cite{Fontes:2014tga,Ferreira:2014dya,Biswas:2015zgk,Ferreira:2014naa}.

\begin{equation}
\cos(\beta -\alpha)=2\cot\beta \hspace{2em}\text{with} \hspace{0.9em}\tan\beta>>2.
\end{equation}
This makes the down-type VLQ couplings with the nonstandard neutral scalar large (they are enhanced by $\tan\beta$), and for $\mathcal{B}_1$ or $\mathcal{B}_2$, the coupling can easily be non-perturbative. We will keep our analysis confined to not-too-large values of $\tan\beta$,and the mass splitting between $\mathcal{B}_1$ and $\mathcal{B}_2$ with a tacit understanding that any other issues with the stability of VLQs are somehow taken care of, most probably by some other dynamics. 

\section{Constraints Imposed on the Parameter Space} \label{constraints}

Before delving into our analysis of how VLQs impact the magnetic moment of muons, denoted as $\Delta a_{\mu}$, it is pertinent to provide a brief overview of the constraints that govern the permissible parameter space. These constraints emanate from a variety of sources, including experimental limitations on particle masses, precision electroweak observations, and theoretical considerations. Experimental constraints on particle masses are essential for ensuring the model's consistency with real-world observations. This includes setting bounds on the masses of particles within the theoretical framework.
Moreover,  electroweak precision data play a crucial role in validating the model's precision. They involve scrutinizing the interactions between particles, particularly those within the electroweak sector, to ensure that the model's predictions align with empirical measurements.
Lastly, theoretical considerations guide the formulation of the model and its parameters. These considerations encompass principles of symmetry, consistency, and the avoidance of phenomena such as flavor-changing neutral currents.

Taken together, these constraints provide a robust foundation for our subsequent analysis, where we investigate how the inclusion of VLQs influences the $(g - 2)_{\mu}$, a fundamental property with significant implications for particle physics.

\subsection{Perturbativity, Vacuum Stability, and Unitarity}

To guarantee the perturbative behavior of all quartic couplings, we establish a critical condition: $\lambda_i \leq 4\pi$. This condition is essential to prevent any of the couplings from reaching values that could lead to non-perturbative behavior.

Additionally, it is imperative to safeguard against scenarios where the potential becomes infinitely negative along any direction in the field space. This is achieved by enforcing conditions that ensure the potential remains bounded from below, which is a crucial requirement for model stability. These necessary and sufficient conditions for ensuring the potential's lower bound have been outlined in \cite{Das:2015mwa}.

In the end, we imposed constraints on the new parameters based on unitarity bounds. Unitarity conditions for 2HDM parameters are expressed in terms of S-matrix eigenvalues, as outlined in  \cite{Ginzburg:2005dt, Horejsi:2005da}. It is necessary to ensure that the following conditions are met to satisfy the unitarity bound: $|\mathsf{a}_i^{\pm}|$ and $|\mathsf{b}_i^{\pm}| \leq 16\pi$ where $\mathsf{a}_i$ and $\mathsf{b}_i$ represent the S-matrix eigenvalues.

\subsection{Constraints from direct searches at the LHC}

Comprehensive searches for VLQs were carried out at the Tevatron \cite{CDF:2011bsz,CDF:2011buy} and LHC \cite{ATLAS:2011mda,CMS:2012dwa}. But no signs of the existence of quarks have been found other than those  in the SM .  These searches have established lower bounds on the mass of VLQs, which are contingent on the specific assumptions regarding their decay modes.

For instance, in scenarios where the \textsc{VLQs} $\mathcal{T}$  or $\mathcal{B}$  solely decay into specific final states such as $Zt/Wb/ht$ or $Zb/Wt/hb$, the mass bounds are exceptionally stringent, with $M_\mathcal{T} > 1310$ GeV and $M_\mathcal{B} < 1030$ GeV \cite{ATLAS:2018ziw}. These bounds can be somewhat relaxed by permitting other decay channels for \textsc{VLQs} \cite{Chala:2017xgc}. For the $\mathcal{T}$ or $\mathcal{B}$ quarks decaying in $\mathsf{q}$ and being associated with $W^{\pm}$ and $Z$ bosons, the VLQ mass bound stands at $M_{\mathcal{T,B}} > 690$. Furthermore, if the $H^{\pm}q$ decay mode is also accessible, the lower limit on VLQ masses may be even less stringent.
In the case of Vector-Like Leptons (VLLs), extensive searches for multi-leptonic events at the LHC have been conducted. These searches have yielded mass bounds, with the ATLAS data suggesting $M_L$ exceeding 300 GeV, while the CMS data indicates $M_L$ greater than 270 GeV \cite{Dermisek:2014qca, Falkowski:2013jya}. These results provide essential insights into the mass ranges of VLQs and VLLs based on the available experimental data.
For the numerical analysis, therefore, we consider  and $M_Q > 1$ TeV for the VLQs. 

\subsection{Effects of VLQs on $h-\gamma-\gamma$ and $h-g-g$}

The single-family \textsc{VLQs} includes a pair of up-type fermions designated as $\mathcal{T}_{1}$ and $\mathcal{T}_{2}$, along with two down-type fermions, $\mathcal{B}_{1}$ and $\mathcal{B}_{2}$. This specific arrangement of new fermions provides a crucial feature: it enables substantial cancellation effects among various contributions from \textsc{VLQs} concerning Higgs precision data and electroweak oblique parameters. The constraints placed on the masses of \textsc{VLQs} by collider experiments, as discussed in the preceding section, primarily impact the three-body decay modes of the Higgs boson. Nevertheless, the change has a minor impact on the overall decay width of Higgs boson.  Charged VLQs have the potential to modify the production of Higgs bosons and their subsequent decay into photons ($\gamma \gamma$). Additionally, the charged Higgs bosons originating from the 2HDM can influence the Higgs 2-photon decay. The $hgg$ and $h\gamma\gamma$ vertices are constrained by the present Higgs precision measurement. The vertices can be altered by the new VLQs. For $m_h = 126$ GeV, the loop functions found in the following vertices, $F_{1}(x_W) \rightarrow + 8.3$  and $F_{1/2}(x_t) \rightarrow-1.34$ V reflect constant values.  One way to accomplish this is to examine the signal strength on that particular channel, which is denoted as $\mu_{\gamma \gamma}$. The present experimental bound is $\mathsf{\mu}_{\gamma\gamma}\equiv\frac{{\mathsf{\mu}^{exp}{\gamma \gamma}}}{{\mathsf{\mu}^{SM}{\gamma \gamma}}}= 1.18^{+0.17}_{-0.14}$ \cite {CMS:2018piu}. Hence, it is imperative to ensure that our model aligns with the current constraints on the Higgs-to-diphoton decay.

The Higgs production can be determined from the Higgs decay width into gluons, and the diphoton decay width can be expressed in terms of the couplings to the particles involved in the loop, as follows:

\begin{equation}
\begin{split}
\Gamma_{h\rightarrow \gamma \gamma} &=\frac{G_{F} \alpha^2 m_h^3}{128\sqrt{2}\pi^3}\Bigg| \kappa_V \mathsf{F}_1(x_W)+   \frac{4}{3} \mathsf{F}_{1/2}(x_t)+\sum_{\mathcal{F}} \sum_{i} N_c Q_\mathcal{F}^2 y^h_{\mathcal{F}_i \mathcal{F}_i} \mathsf{F}_{1/2}(x_{\mathcal{F}_i})\\&+  \mathsf{k}_{\mathsf{h} \mathsf{H}^{+} \mathsf{H}^{-}} \mathsf{F}_{+}(x_{H^{\pm}})\Bigg|^2 
\end{split}
 \end{equation}

\begin{equation}
\begin{split}
\Gamma_{h\rightarrow g g}& =\frac{G_{F} \alpha^2 m_h^3}{64\sqrt{2}\pi^3}\Bigg| \mathsf{F}_{1/2}(x_t) + \sum_{\mathcal{F}} \sum_{i}  y^h_{\mathcal{F}_i \mathcal{F}_i} \mathsf{F}_{1/2}(x_{\mathcal{F}_i})\Bigg|^2 	
 \end{split}
 \end{equation}

where $\mathcal{F} = \mathcal{T}, \mathcal{B}$ , i = 1, 2$x_j = (\frac{2M_j} { m_h })^2 , (j = W,t,\mathcal{F}, H^{\pm} $). The corresponding  loop functions $\mathsf{F}_1 , \mathsf{F}_{1/2}$ and $\mathsf{F}_+$ which appear in the calculation are 
\begin{equation} \label{eq4.26}
\begin{split}
\mathsf{F}_1(x)&=2 + 3 x +3 x (2-x)f(x) \\  
\mathsf{F}_{1/2}(x)&=-2x[ 1+ (1-x)f(x)] \\  
\mathsf{F}_{+}(x)&=-x[1-x f(x)]
\end{split}
\end{equation}
\[
f(x)= 
\begin{cases}
(\sin^{-1} (1/\sqrt{x}))^2,&  x\geq 1\\
-\frac{1}{4}(\ln(\frac{1+\sqrt{1-x}}{1-\sqrt{1-x}})-i \pi)^2,  & x<1
\end{cases}
\]
and, the charged Higgs couplings to the SM Higgs is given by \cite{Djouadi:1996yq}

\begin{equation}
	\mathsf{k}_{\mathsf{h}\mathsf{H}^+ \mathsf{H}^-}=-\frac{1}{2M_{\mathsf{H}^{\pm}}^2} \Big[\frac{(m_\mathsf{h}^2 - 2M_{\mathsf{H}^{\pm}}^2)\cos(\alpha- 3\beta)+ (3m_\mathsf{h}^2 + 2M_{\mathsf{H}^{\pm}}^2 -4 m_0^2)\cos(\alpha + \beta)}{4\sqrt{2}  \sin\beta\cos\beta}\Big]
\end{equation}
where
\begin{equation}
	m_0^2 = \frac{m_{12}^2}{\sin\beta \cos\beta}
\end{equation}

\begin{figure}[htp!]
	\centering
\subfloat[]	{\includegraphics[width=0.48\textwidth]{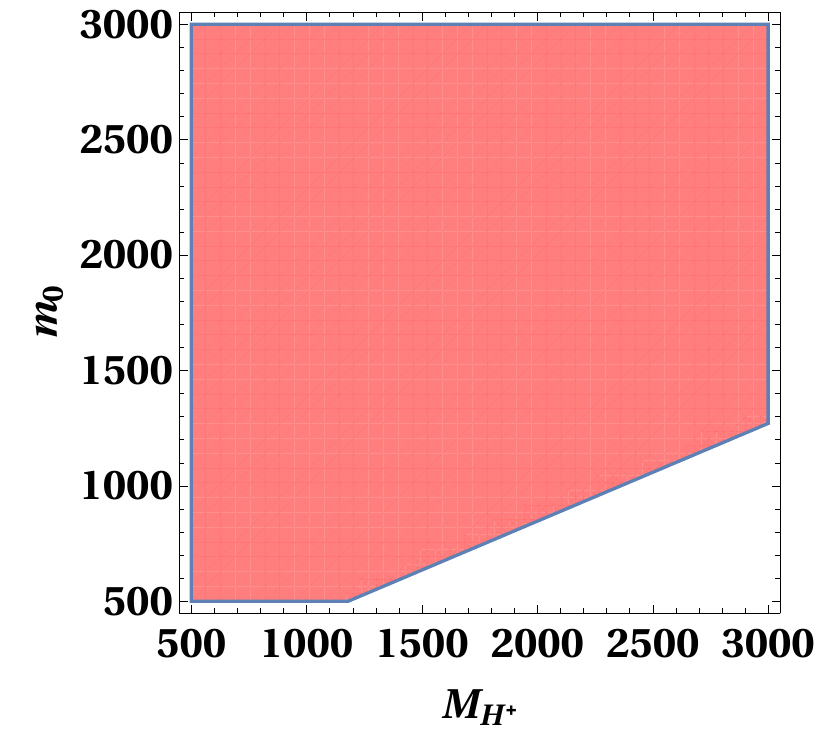}}\label{fig:A2}
\subfloat[]	{\includegraphics[width=0.48\textwidth]{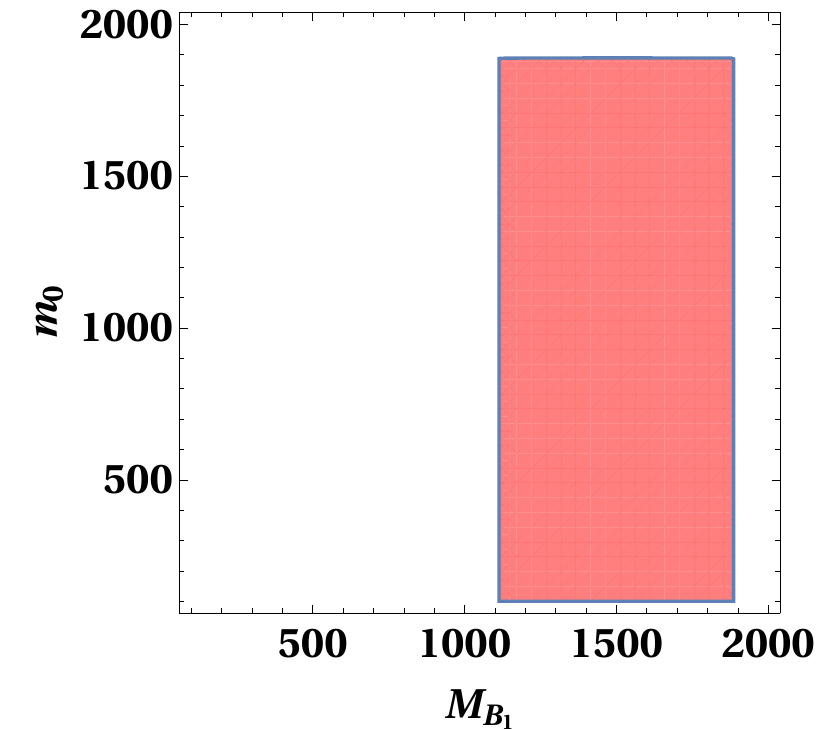}}
\begin{figure}[H]
	\centering
\subfloat[]	{\includegraphics[width=0.48\textwidth]{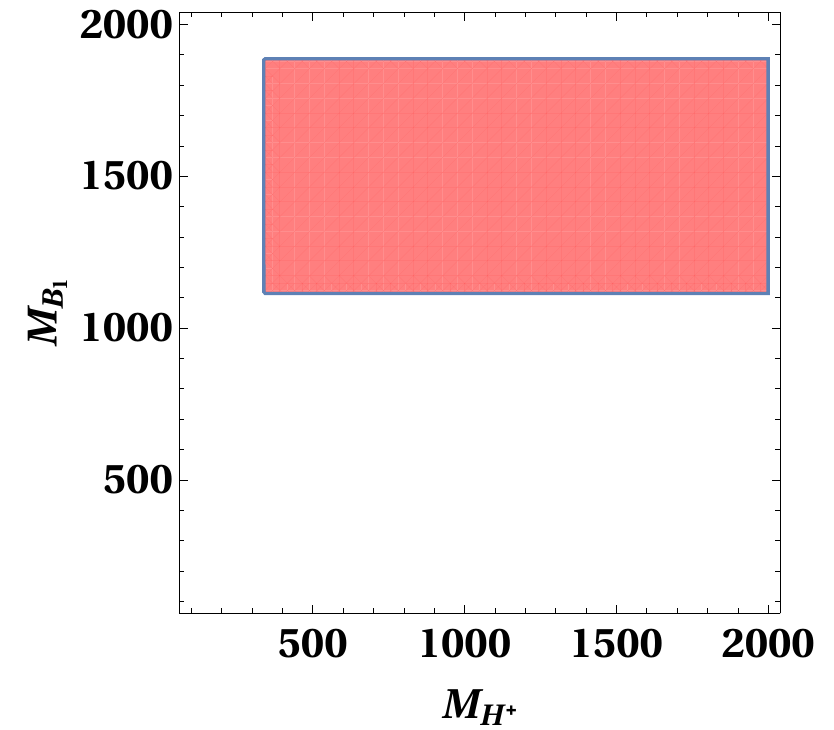}}\label{fig:A1}
\subfloat[]	{\includegraphics[width=0.48\textwidth]{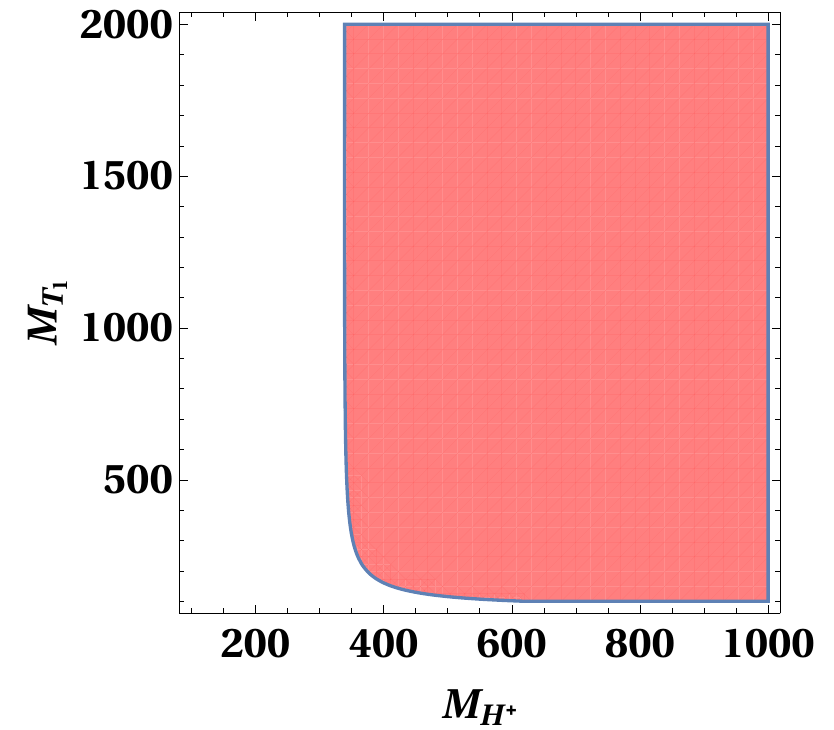}}\label{fig:B1}
	\label{fig:tanbmlme} 
\end{figure}

\caption{{\emph{ The $2\sigma$   allowed region from $\mu_{\gamma\gamma}$ signal strength in (a) ($ M_{H^{\pm}} - m_0$ ) , (b) ($ M_{\mathcal{B}_1}-m_0$ ), (c)($M_{H^{\pm}} - M_{\mathcal{B}_1}$ ) and (d)($M_{H^{\pm}} - M_{\mathcal{T}_1}$ ) planes.}}}
	\label{fig:hgamma} 
\end{figure}

In Fig.\eqref{fig:hgamma}, we present a graphical representation of the constraints on various mass parameters, including $m_0$, $M_{H^{\pm}}$, $M_{\mathcal{B}_1}$, and $M_{\mathcal{T}_1}$. These constraints are applied under the condition that the Yukawa couplings $|Y_{\mathcal{B}}| < 4 \pi$, with a fixed value of $\tan\beta$ ($\tan\beta=12$), and specific choices for the \textsc{VLQs} mixing angles ($\theta_{\mathcal{B}} = 0.96$ and $\theta_{\mathcal{T}} = 0.7$). 

Figure \eqref{fig:hgamma}(a) shows that, with a constant VLQ mass and a specific value assigned to $M_{H^{\pm}}$, there exists a significant range within which the mass parameter $m_0$ is permissible.The depiction clearly suggests that we can confidently select the scenario where $m_0$ is equal to $M_{H^{\pm}}$ for our analysis.

 The allowed region in Fig.\eqref{fig:hgamma}(b) shows that for a specific choice of $M_{\mathcal{B}_1}$ mass a large range of $m_0$ is  allowed.  Also in Fig.\eqref{fig:hgamma}(c) we can see that for a fixed value of $\mathcal{B}_1$ a large range of $M_{H^{\pm}}$ allowed. But for $ M_{H^{\pm}} - M_{\mathcal{T}_1} $ plane its look quite different for a fixed $M_{H^{\pm}}$ value more range of $M_{\mathcal{T}_1}$ allowed than $M_{\mathcal{B}_1}$. This is happening due to the Yukawa perturbative restriction $|Y_{\mathcal{T},\mathcal{ B}}|< 4 \pi$.

Based on the analysis provided, we can draw the conclusion that there exists a favorable region within the model parameter space from the constraints imposed by $\mathsf{\mu}_{\mathsf{\gamma} \mathsf{ \gamma}}$.

\subsection{Constraints arising from oblique  parameters $\mathrm{S,T,U}$}

In addition to the constraint we have previously mentioned, we need to take into account the restrictions originating from the parameters $S$, $T$, and $U$. This is because loop corrections allow the VLQs to contribute to the masses of gauge bosons.

The 2HDM scalar's impact on the $S$, $T$, and $U$ parameters is significant when integrated with VLQs. The formulation of the scalar component has been thoroughly explored and documented in earlier work in \cite{Grimus:2007if, Grimus:2008nb}. In order to incorporate the effects of \textsc{VLQs}, we begin by computing the interactions between \textsc{VLQs} and gauge bosons. The couplings of \textsc{VLQs} with $W$ and $Z$ bosons can be represented by the following expressions:

\begin{equation}
\begin{split}
\mathcal{L}^Z &= \frac{g}{2\cos\theta_{\mathsf{W}}}\Big(\bar{\mathcal{Q}}_{L}\gamma^{\mu} c_V^{F'} \mathcal{Q}_{L} + \bar{\mathcal{Q}}_{R}\gamma^{\mu} c_V^{F'} \mathcal{Q}_{R}+  \bar{\mathcal{B}}_{L}\gamma^{\mu} c_V^{F} \mathcal{B}_{L}  \\& +  \bar{\mathcal{B}}_{R}\gamma^{\mu} c_V^{F} \mathcal{B}_{R}   +  \bar{\mathcal{T}}_{L}\gamma^{\mu} c_V^{F} \mathcal{T}_{L}  +  \bar{\mathcal{T}}_{R}\gamma^{\mu} c_V^{F} \mathcal{T}_{R}    \Big)Z_{\mu} \\
&=\frac{g}{2\cos\theta_{\mathsf{W}}}\sum_{i,j=1}^{2}\Big( g^{\mathsf{Z\mathcal{B}_{i} \mathcal{B}_{j}}}_{L}\bar{\mathcal{B}}_{iL}\gamma^{\mu}  \mathcal{B}_{jL}   +   g^{\mathsf{Z\mathcal{B}_{i} \mathcal{B}_{j}}}_{R} \bar{\mathcal{B}}_{iR}\gamma^{\mu}  \mathcal{B}_{jR}   \\& +  g^{\mathsf{Z\mathcal{T}_{i} \mathcal{T}_{j}}}_{L} \bar{\mathcal{T}}_{iL}\gamma^{\mu}  \mathcal{T}_{jL}  +  g^{\mathsf{Z\mathcal{T}_{i} \mathcal{T}_{j}}}_{R}  \bar{\mathcal{T}}_{iR}\gamma^{\mu}  \mathcal{T}_{jR}    \Big)Z_{\mu}
\end{split}
\end{equation}

\begin{equation}
\begin{split}
\mathcal{L}^W &= \frac{g}{\sqrt{2}}\Big(\bar{\mathcal{B}}'_{L}\gamma^{\mu}  \mathcal{T}'_{L} + \bar{\mathcal{B}'}_{R}\gamma^{\mu}  \mathcal{T}'_{R}+  \bar{\mathcal{B}'}_{L}\gamma^{\mu}  \mathcal{T}_{L}  +  \bar{\mathcal{B}'}_{R}\gamma^{\mu}  \mathcal{T}_{R}   \\& +  \bar{\mathcal{B}}_{L}\gamma^{\mu}  \mathcal{T}'_{L}  +  \bar{\mathcal{B}}_{R}\gamma^{\mu}  \mathcal{T}'_{R}  +  \bar{\mathcal{B}}_{L}\gamma^{\mu}  \mathcal{T}_{L}  +  \bar{\mathcal{B}}_{R}\gamma^{\mu}  \mathcal{T}_{R}  \Big) W^+_{\mu}   + \mathsf{H.c}\\
&=\frac{g}{\sqrt{2}}\sum_{i,j=1}^{2}  \Big( \Big[ g^{\mathsf{W\mathcal{B}_{i} \mathcal{T}_{j}}}_{L}\bar{\mathcal{B}}_{iL}\gamma^{\mu}  \mathcal{T}_{jL}   +   g^{\mathsf{W\mathcal{B}_{i} \mathcal{T}_{j}}}_{R} \bar{\mathcal{B}}_{iR}\gamma^{\mu}  \mathcal{T}_{jR}      \Big] \Big)  W^+_{\mu} + \mathsf{H.c}
\end{split}
\end{equation}

We have outlined the equations for the incorporation of VLQs contributions into the $S$ and $T$ parameters \cite{Anastasiou:2009rv,Lavoura:1992np,ATLAS:2015uhg,Chen:2017hak}. It is important to note that in our numerical analysis, we also account for the influence of the $U$ parameter. Thus, our comprehensive analysis covers all three parameters, namely, $S$, $T$, and $U$.

\begin{equation}
\begin{split}
\Delta \mathsf{T}_\mathcal{F} &=\frac{1}{8 \pi s_W^2 c_W^2}\sum_{i,j}\Big[\big(|g^{\mathsf{W\mathcal{B}_{i} \mathcal{T}_{j}}}_{L}|^2 + |g^{\mathsf{W\mathcal{B}_{i} \mathcal{T}_{j}}}_{R}|^2\big) \mathrm{\theta}_{+}(\mathcal{F}_a,\mathcal{F}_b) + 2 \mathsf{Re}\big(g^{\mathsf{W\mathcal{B}_{i} \mathcal{T}_{j}}}_{L}  g^{\mathsf{W\mathcal{B}_{i} \mathcal{T}_{j}}}_{R}\big) \theta_{-}(\mathcal{F}_a,\mathcal{F}_b)  \\& -\frac{1}{2} \big(|g^{\mathsf{Z\mathcal{B}_{i} \mathcal{B}_{j}}}_{L}|^2 + |g^{\mathsf{Z\mathcal{B}_{i} \mathcal{B}_{j}}}_{R}|^2\big) \theta_{+}(\mathcal{F}_a,\mathcal{F}_b) + 2\mathsf{Re}\big(g^{\mathsf{Z\mathcal{B}_{i} \mathcal{B}_{j}}}_{L}  g^{\mathsf{Z\mathcal{B}_{i} \mathcal{B}_{j}}}_{R}\big) \theta_{-}(\mathcal{F}_a,\mathcal{F}_b) \Big]
\end{split}
\end{equation}

\begin{equation}
\begin{split}
\Delta \mathsf{S}_\mathcal{F} &=\frac{1}{2 \pi}\sum_{ij}\Big[\big(|g^{\mathsf{W\mathcal{B}_{i} \mathcal{T}_{j}}}_{L}|^2 + |g^{\mathsf{W\mathcal{B}_{i} \mathcal{T}_{j}}}_{R}|^2\big) \psi_{+}(\mathcal{F}_a,\mathcal{F}_b) + 2 \mathsf{Re}\big(g^{\mathsf{W\mathcal{B}_{i} \mathcal{T}_{j}}}_{L}  g^{\mathsf{W\mathcal{F}_{i} \mathcal{T}_{j}}}_{R}\big) \psi_{-}(\mathcal{F}_a,\mathcal{F}_b)    \\& -\frac{1}{2} \big(|g^{\mathsf{Z\mathcal{B}_{i} \mathcal{B}_{j}}}_{L}|^2 + |g^{\mathsf{Z\mathcal{B}_{i} \mathcal{B}_{j}}}_{R}|^2\big) \xi_{+}(\mathcal{F}_a,\mathcal{F}_b) + 2 \mathsf{Re}\big(g^{\mathsf{Z\mathcal{B}_{i} \mathcal{B}_{j}}}_{L} g^{\mathsf{Z\mathcal{B}_{i} \mathcal{B}_{j}}}_{R}\big) \xi_{-}(\mathcal{F}_a,\mathcal{F}_b) \Big]
\end{split}
\end{equation}

 where  $\mathcal{F}_a =\frac{M_{\mathcal{F}_{a}}^2}{M_Z^2}$ and the functions are defined as

\begin{eqnarray}
\psi_{+}(x,y) = \frac{1}{3} -\frac{1}{9}\ln(\frac{x}{y})  \nn \\ 
\psi_{-}(x,y) = -\frac{x + y}{6 \sqrt{x y}}  \nn \\
\theta_{+}(x,y)= 
\begin{cases}
	\frac{x+y}{2}-\frac{x y}{x-y} \ln(\frac{x}{y}),& \text{if } x\neq y \\
	0,              & \text{if } x = y
\end{cases}
\nn \\
\theta_{-}(x,y)= 
\begin{cases}
	\sqrt{x y} \Big[  \frac{x+y}{x -y} \ln(\frac{x}{y}) -2  \Big],& \text{if } x\neq y \\
	0,              & \text{if } x = y
\end{cases}
\nn \\
\xi_{+}(x,y)= 
\begin{cases}
	\frac{5 (x^2 + y^2) - 22 x y }{9 (x- y)^2} + \frac{ 3 x y (x + y ) - x^3 - y^3 }{3 (x-y)^3} \ln(\frac{x}{y}),& \text{if } x\neq y \\
	0,              & \text{if } x = y
\end{cases}
\nn \\
\xi_{-}(x,y)= 
\begin{cases}
	\sqrt{x y} \Big[  \frac{x+y}{6 x y}  - \frac{x + y }{(x - y)^2} + \frac{2 x y }{(x - y)^3}\ln(\frac{x}{y})  \Big],& \text{if } x\neq y \\
	0,              & \text{if } x = y
\end{cases}
\end{eqnarray}  \label{eq4.32}

\begin{table}[H]
\begin{center}
	\begin{tabular}{ |c|c||c|c| } 
		\hline
	\textit{Coupling with} $Z$ & \textit{Expression} & \textit{Coupling with} $W^{\pm}$  &  \textit{Expression}   \\ 
		\hline
	    $g^{\mathsf{Z\mathcal{B}_{1} \mathcal{B}_{1}}}_{L} $       & $c_{V}^{\mathcal{B}'} \cos^2 \theta^{\mathcal{B}}_L  +  c_{V}^{\mathcal{B}} \sin^2 \theta^{\mathcal{B}}_L$   &  $g^{\mathsf{W\mathcal{B}_{1} \mathcal{T}_{1}}}_{L}$   &    $(\cos\theta^{\mathcal{T}}_L - \sin\theta^{\mathcal{T}}_L) (\cos\theta^{\mathcal{B}}_L - \sin\theta^{\mathcal{B}}_L)$\\ 
      \hline
	    $g^{\mathsf{Z\mathcal{B}_{2} \mathcal{B}_{2}}}_{L} $             & $c_{V}^{\mathcal{B}'} \sin^2 \theta^{\mathcal{B}}_L  +  c_{V}^{\mathcal{B}} \cos^2 \theta^{\mathcal{B}}_L$    & $g^{\mathsf{W\mathcal{B}_{1} \mathcal{T}_{2}}}_{L}$    &     $(\cos\theta^{\mathcal{B}}_L - \sin\theta^{\mathcal{B}}_L) (\sin\theta^{\mathcal{T}}_L + \cos\theta^{\mathcal{T}}_L)$\\ 
      \hline
	     $g^{\mathsf{Z\mathcal{B}_{1} \mathcal{B}_{2}}}_{L} = g^{\mathsf{Z\mathcal{B}_{2} \mathcal{B}_{1}}}_{L} $                      & $(c_{V}^{\mathcal{B}'} - c_{V}^{\mathcal{B}})\cos \theta^{\mathcal{B}}_L \sin \theta^{\mathcal{B}}_L$ &  $g^{\mathsf{W\mathcal{B}_{2} \mathcal{T}_{1}}}_{L}$    &    $(\sin\theta^{\mathcal{B}}_L + \cos\theta^{\mathcal{B}}_L) (\cos\theta^{\mathcal{T}}_L - \sin\theta^{\mathcal{T}}_L)$\\ 
      \hline
	    $g^{\mathsf{Z\mathcal{T}_{1} \mathcal{T}_{1}}}_{L} $         & $c_{V}^{\mathcal{T}'} \cos^2 \theta^{\mathcal{T}}_L  +  c_{V}^{\mathcal{T}} \sin^2 \theta^{\mathcal{T}}_L$   &  $g^{\mathsf{W\mathcal{B}_{2} \mathcal{T}_{2}}}_{L}$    &   $(\sin\theta^{\mathcal{B}}_L + \cos\theta^{\mathcal{B}}_L) (\sin\theta^{\mathcal{T}}_L + \cos\theta^{\mathcal{T}}_L)$ \\ 
      \hline
	      $g^{\mathsf{Z\mathcal{T}_{2} \mathcal{T}_{2}}}_{L} $          & $c_{V}^{\mathcal{T}'} \sin^2 \theta^{\mathcal{T}}_L  +  c_{V}^{\mathcal{T}} \cos^2 \theta^{\mathcal{T}}_L$   &   $g^{\mathsf{W\mathcal{B}_{1} \mathcal{T}_{1}}}_{R}$     & $(\cos\theta^{\mathcal{T}}_R - \sin\theta^{\mathcal{T}}_R) (\cos\theta^{\mathcal{B}}_R - \sin\theta^{\mathcal{B}}_R)$\\ 
      \hline
	      $g^{\mathsf{Z\mathcal{T}_{1} \mathcal{T}_{2}}}_{L} = g^{\mathsf{Z\mathcal{T}_{2} \mathcal{T}_{1}}}_{L} $      & $(c_{V}^{\mathcal{T}'} - c_{V}^{\mathcal{T}})\cos \theta^{\mathcal{T}}_L \sin \theta^{\mathcal{T}}_L$  &  $g^{\mathsf{W\mathcal{B}_{1} \mathcal{T}_{2}}}_{R}$    & $(\cos\theta^{\mathcal{B}}_R - \sin\theta^{\mathcal{B}}_R) (\sin\theta^{\mathcal{T}}_R + \cos\theta^{\mathcal{T}}_R)$\\ 
      \hline
	    $g^{\mathsf{Z\mathcal{B}_{1} \mathcal{B}_{1}}}_{R} $       & $c_{V}^{\mathcal{B}'} \cos^2 \theta^{\mathcal{B}}_R  +  c_{V}^{\mathcal{B}} \sin^2 \theta^{\mathcal{B}}_R$  &    $g^{\mathsf{W\mathcal{B}_{2} \mathcal{T}_{1}}}_{R}$  & $(\sin\theta^{\mathcal{B}}_R + \cos\theta^{\mathcal{B}}_R) (\cos\theta^{\mathcal{T}}_R - \sin\theta^{\mathcal{T}}_R)$\\ 
      \hline
	    $g^{\mathsf{Z\mathcal{B}_{2} \mathcal{B}_{2}}}_{R} $             & $c_{V}^{\mathcal{B}'} \sin^2 \theta^{\mathcal{B}}_R  +  c_{V}^{\mathcal{B}} \cos^2 \theta^{\mathcal{B}}_R$  &   $g^{\mathsf{W\mathcal{B}_{2} \mathcal{T}_{2}}}_{R}$   &  $(\sin\theta^{\mathcal{B}}_R + \cos\theta^{\mathcal{B}}_R) (\sin\theta^{\mathcal{T}}_R + \cos\theta^{\mathcal{T}}_R)$\\ 
      \hline
	     $g^{\mathsf{Z\mathcal{B}_{1} \mathcal{B}_{2}}}_{R} =g^{\mathsf{Z\mathcal{B}_{2} \mathcal{B}_{1}}}_{R} $            & $(c_{V}^{\mathcal{B}'} - c_{V}^{\mathcal{B}})\cos \theta^{\mathcal{B}}_R \sin \theta^{\mathcal{B}}_R$  &     &\\ 
      \hline
	    $g^{\mathsf{Z\mathcal{T}_{1} \mathcal{T}_{1}}}_{R} $         & $c_{V}^{\mathcal{T}'} \cos^2 \theta^{\mathcal{T}}_R  +  c_{V}^{\mathcal{T}} \sin^2 \theta^{\mathcal{T}}_R$  &     & \\ 
      \hline
	      $g^{\mathsf{Z\mathcal{T}_{2} \mathcal{T}_{2}}}_{R} $          & $c_{V}^{\mathcal{T}'} \sin^2 \theta^{\mathcal{T}}_R  +  c_{V}^{\mathcal{T}} \cos^2 \theta^{\mathcal{T}}_R$    &     &  \\ 
      \hline
	      $g^{\mathsf{Z\mathcal{T}_{1} \mathcal{T}_{2}}}_{R} = g^{\mathsf{Z\mathcal{T}_{2} \mathcal{T}_{1}}}_{R}$  & $(c_{V}^{\mathcal{T}'} - c_{V}^{\mathcal{T}})\cos \theta^{\mathcal{T}}_R \sin \theta^{\mathcal{T}}_R$     &     &   \\
		\hline
       
	\end{tabular}
\end{center}
\caption{\emph{Required coupling factor of VLQ with $Z$ and $W^{\pm}$ for $S$,$T$,$U$ parameters and their full analytical formulation where  $c^{\mathcal{F}}_V = T^3_\mathcal{F} -Q_\mathcal{F} s_W^2$.}}
\label{tableparam}
\end{table}

The present global electroweak fit provides certain conditions that must be satisfied by the combined contributions of both the scalar particles and the new VLQs. The most intelligent strategy to evade the oblique parameter restrictions within the 2HDM scalar model is to set the masses of the scalar components, namely $M_H$, $M_A$, and $M_{H^{\pm}}$, to be equal. Consequently, in our comprehensive analysis, we have chosen to fix the masses of all scalar particles at 800 GeV.

\begin{figure}[H]
	\centering
\subfloat[]	{\includegraphics[width=0.48\textwidth]{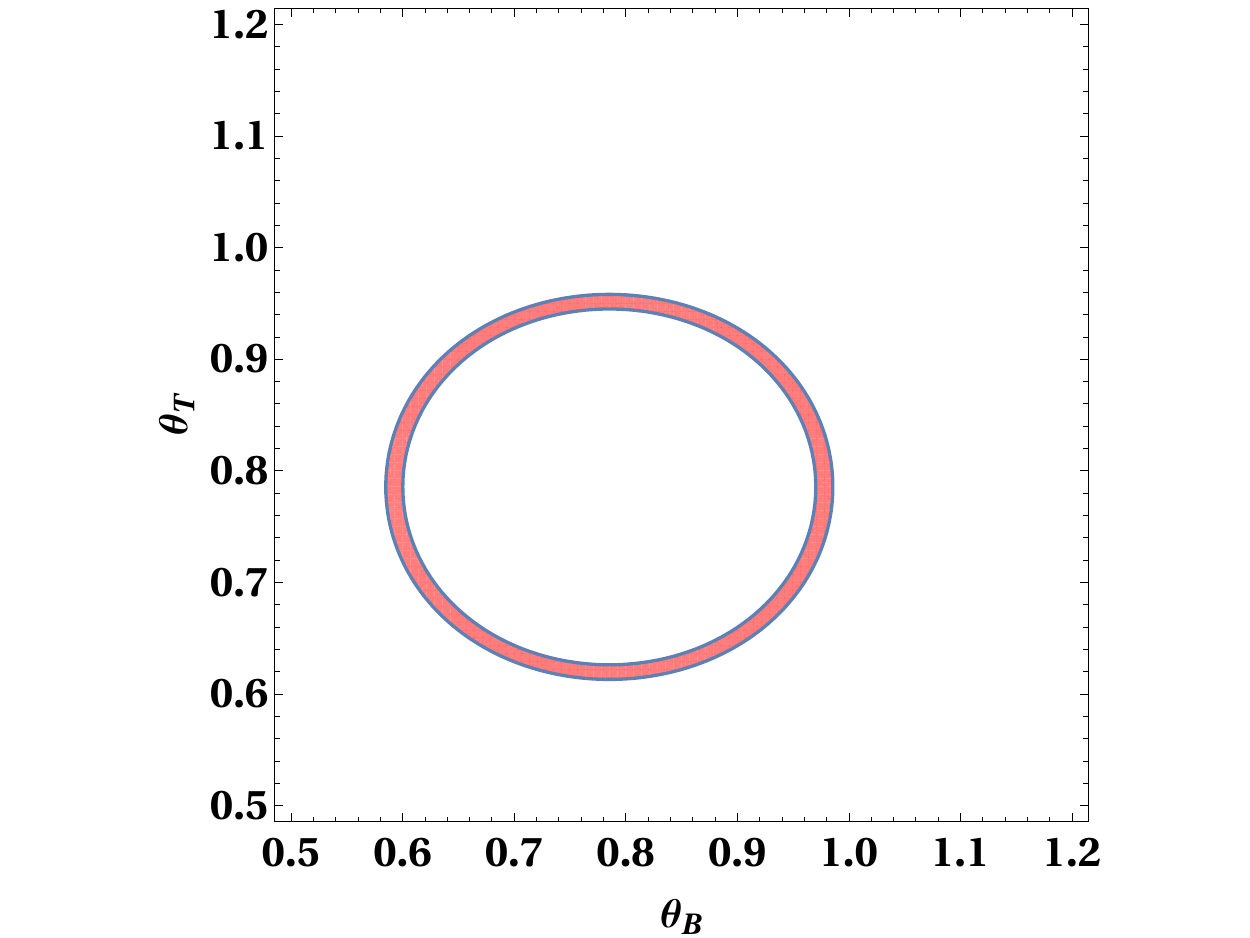}
\subfloat[]	{\includegraphics[width=0.48\textwidth]{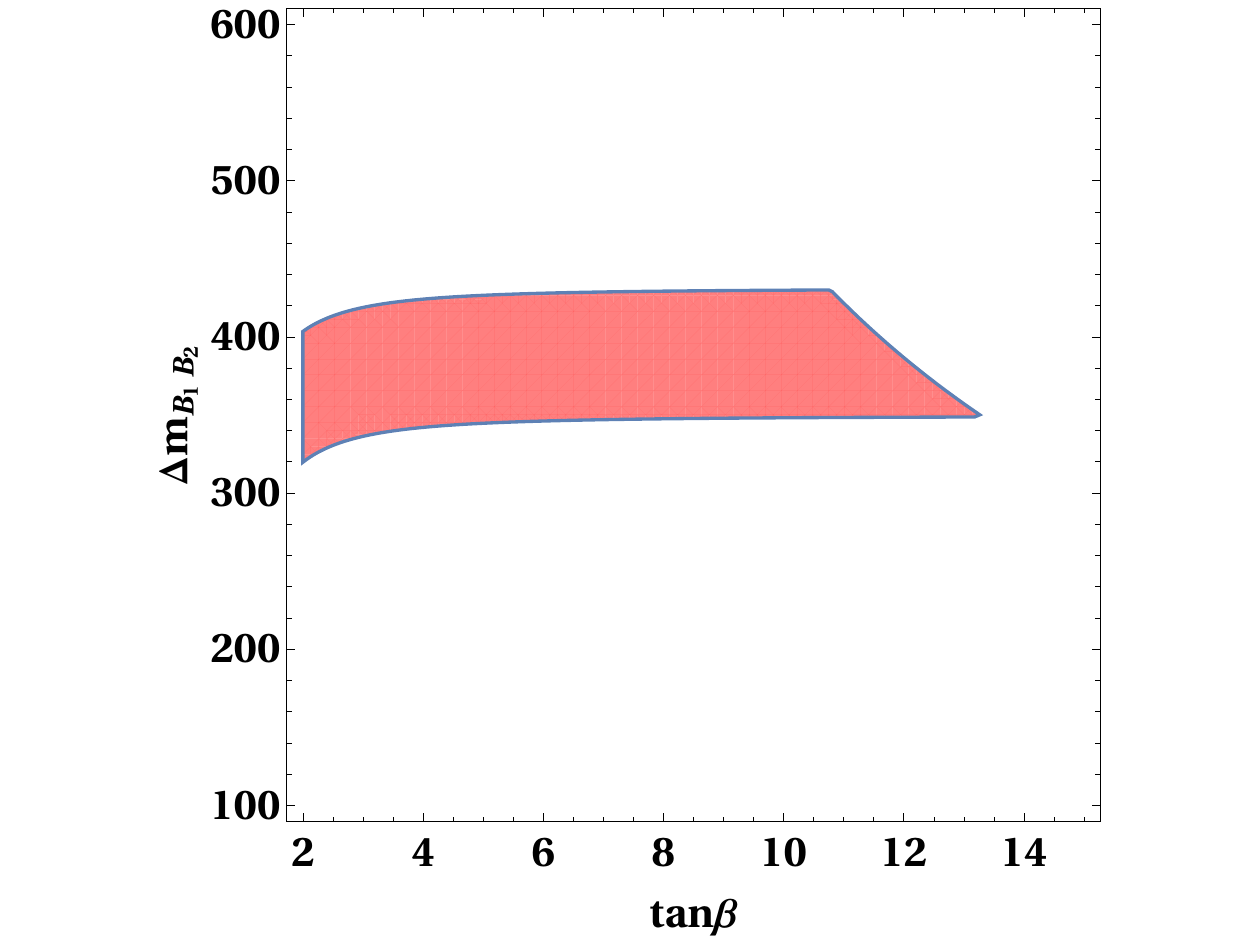}}
\label{fig:A2}}
\caption{Authorized parameter space in (a)$M_{\mathcal{T}_1}-M_{\mathcal{B}_1}$, (b)$M_{\mathcal{T}_2}-M_{\mathcal{B}_2}$, and (c) $\theta_{\mathcal{B}}-\theta_{\mathcal{T}} $ plane comes from $S$,$T$ and $U$ restriction.}
	\label{fig:stu1} 
\end{figure}

In Fig. \eqref{fig:stu1} we have illustrated the constraints on the masses and mixing of VLQs  by $S$, $T$, and $U$ parameters. The constraints are represented in  planes of (a)$M_{\mathcal{T}_1}$-$M_{\mathcal{B}_1}$, (b) $M_{\mathcal{T}_2}$-$M_{\mathcal{B}_2}$, and (c) $\theta_{\mathcal{B}}$-$\theta_{\mathcal{T}}$, taking into account a 2$\sigma$ level of uncertainty.

\begin{figure}[H]

	\centering
\subfloat[]	{\includegraphics[width=0.48\textwidth]{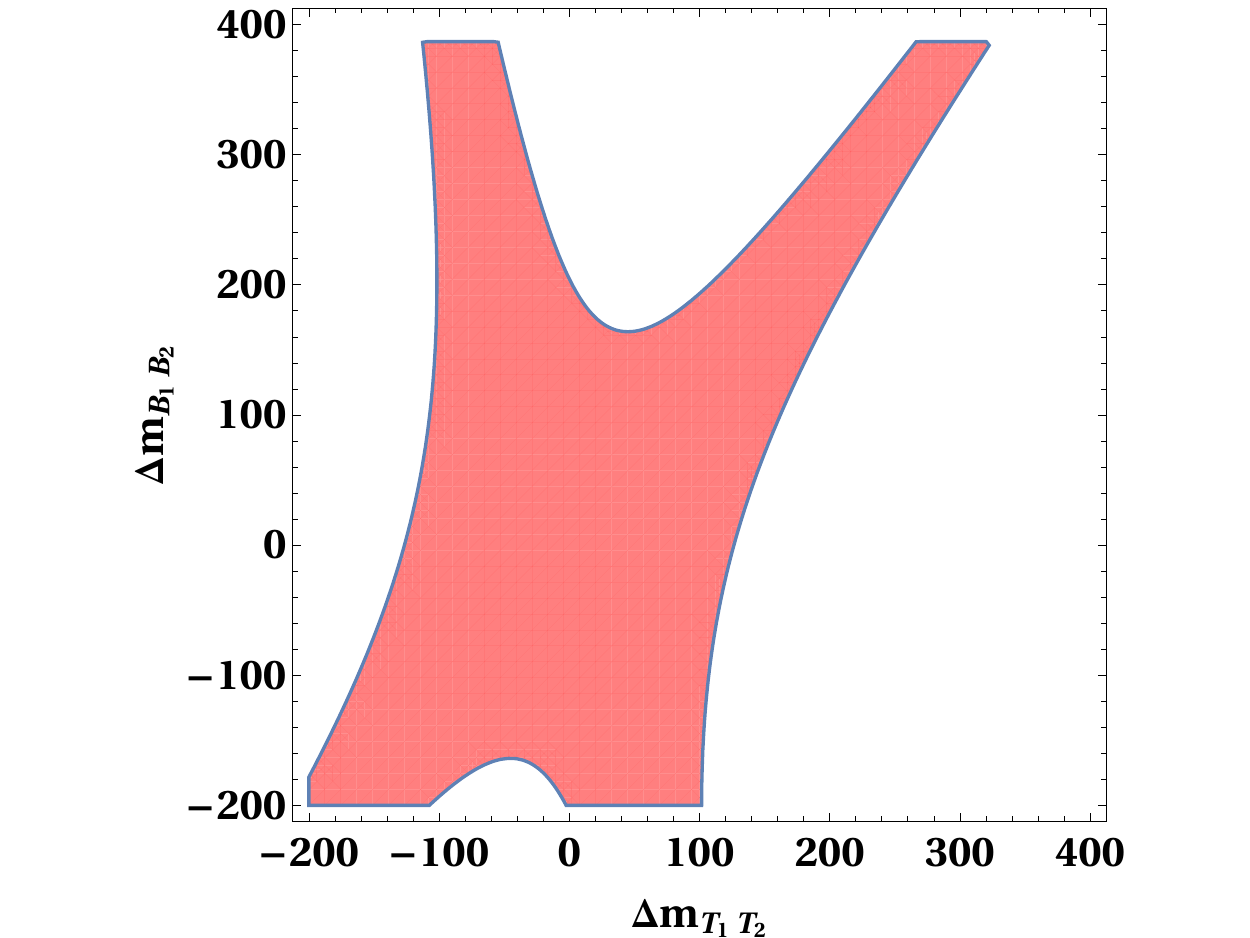}\label{fig:A1}}
\subfloat[]	{\includegraphics[width=0.48\textwidth]{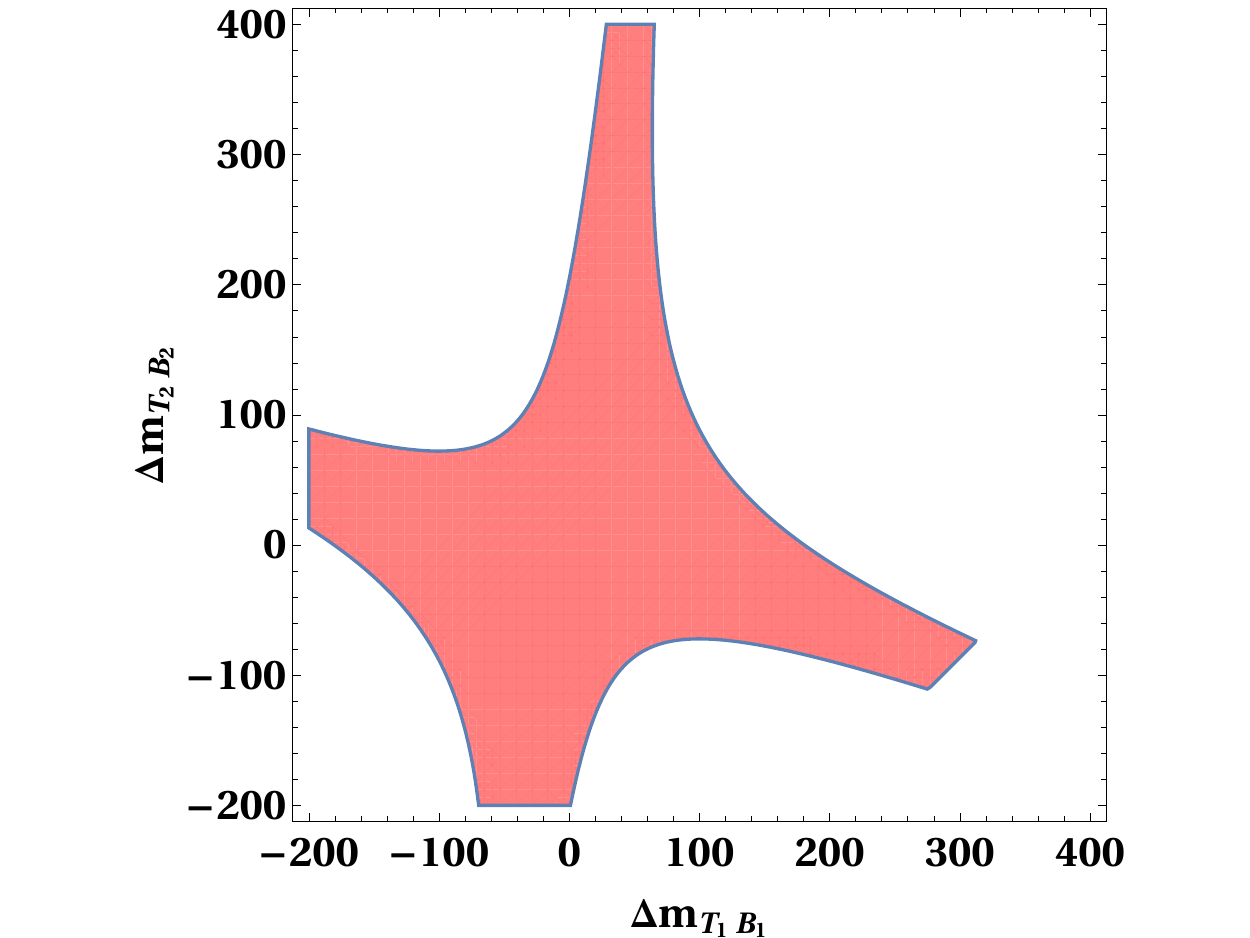}}\label{fig:B1}
\caption{{Authorized parameter space in (a)$\Delta m_{\mathcal{T}_1\mathcal{T}_2}-\Delta m_{\mathcal{B}_1\mathcal{B}_2}$ and(b)$\Delta m_{\mathcal{T}_1\mathcal{B}_1}-\Delta m_{\mathcal{T}_2\mathcal{B}_2}$, plane comes from $S$,$T$ and $U$ restriction.}}
	\label{fig:stu2} 
 
\end{figure}

To gain a deeper comprehension of the correlation in VLQs mass splitting, we have examined the impact of constraints imposed by parameters $S$, $T$, and $U$ on mass splitting. This analysis is presented in two distinct manners in Fig. \eqref{fig:stu2}.

 For the first case Fig. \eqref{fig:stu2}(a) illustrates the mass splitting between $\mathcal{T}_1$ and $\mathcal{T}_2$ (as $\Delta m_{\mathcal{T}_1\mathcal{T}_2}\equiv M_{\mathcal{T}_2}-M_{\mathcal{T}_1}$) and between $\mathcal{B}_1$ and $\mathcal{B}_2$ (as $\Delta m_{\mathcal{B}_1\mathcal{B}_2}\equiv M_{\mathcal{B}_2}-M_{\mathcal{B}_1}$). The graph reveals that achieving a mass splitting of 300 GeV between $\mathcal{T}_{1}$ and $\mathcal{T}_{2}$ necessitates a corresponding mass splitting of 350 GeV between $\mathcal{B}_{1}$ and $\mathcal{B}_{2}$. In the second scenario Fig. \eqref{fig:stu2}(b) presents the mass splitting between $\mathcal{T}_1$ and $\mathcal{B}_1$ (as $\Delta m_{\mathcal{T}_1\mathcal{B}_1}\equiv M_{\mathcal{B}_1}-M_{\mathcal{T}_1}$) and between $\mathcal{T}_2$ and $\mathcal{B}_2$ (as $\Delta m_{\mathcal{T}_2\mathcal{B}_2}\equiv M_{\mathcal{B}_2}-M_{\mathcal{T}_2}$).   With access to this authorized parameter range, we can now probe into how these results can shed light on the $\Delta a_{\mu}$.    
\hspace{0.5cm}
\section{$(g-2)_{\mu}$ with VLQs} \label{muon}

\begin{figure}[htbp!]

	\centering
	%
	\includegraphics[width=1.0\textwidth]{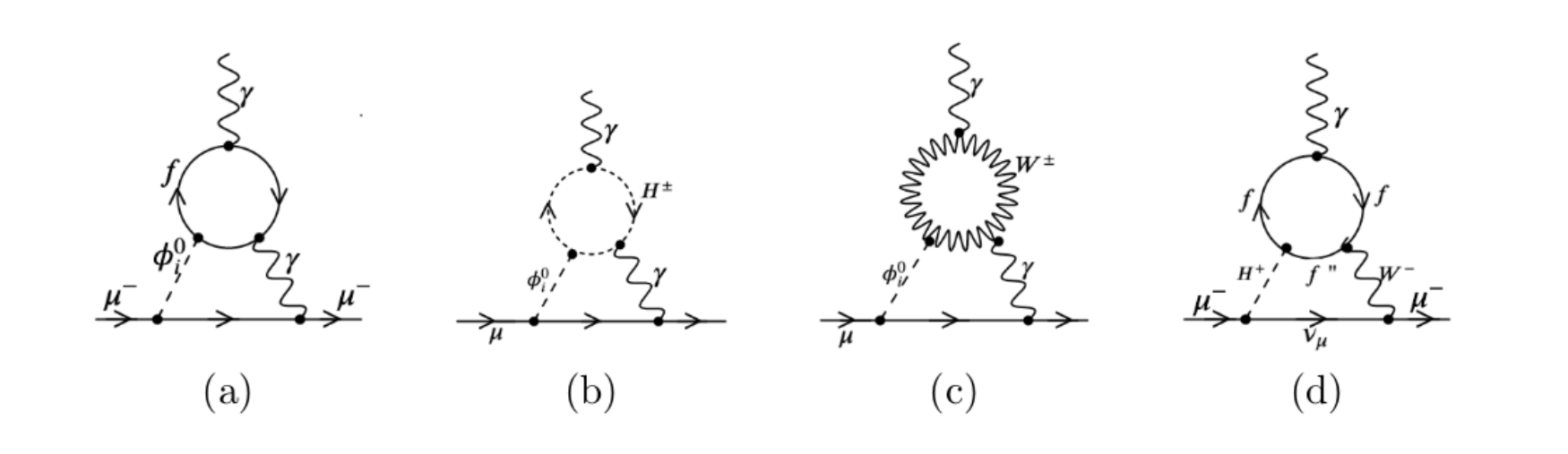}
	\caption{The  2-loop contribution with VLQs where $f,f^{''}= t, b,\mathcal{T}_1,\mathcal{B}_1,\mathcal{T}_2,\mathcal{B}_2$ and $\phi_i^0 = h, H, A$.}
	\label{fig:2loop}
\end{figure}

Let's delve into the primary goal of this research, which is to explore how the incorporation of VLQs within the framework of the X-II 2HDM+VLQs structure contributes to our understanding of $\Delta a_{\mu}$. We will explore the impact of both 1-loop and 2-loop \textsc{Bar-Zee} contributions on $(g-2)_{\mu}$ with VLQs, which has been extensively investigated within the framework of the 2HDM \cite{Jegerlehner:2009ry,Leveille:1977rc,Lynch:2001zs}. In the case of the 2HDM, the 1-loop contribution becomes more extensive when incorporating graphs that involve both $\mathsf{H,A}$ and $\mathsf{H}^{\pm}$ \cite{Abe:2015oca}. The level of enhancement further escalates when we take into account the 2-loop \textsc{B-Z} graphs in the context of the 2HDM.  While the 2-loop contribution is affected by a suppression factor represented by $\frac{\alpha_{EM}}{\pi}$, this limitation can be overcome by introducing heavy particles into the loop, where the factor $\frac{M^2}{m_{\mu}^2}$ plays a crucial role in mitigating the suppression. Reference \cite{Ilisie:2015tra} provides an extensive compilation and a detailed presentation of the analytical results pertaining to all 2-loop contributions.

The diagram that holds the highest significance in the X-II 2HDM+VLQs framework when it comes to the B-Z graph is illustrated in Fig.\eqref{fig:2loop}. Among the various graphs depicted, the (d) graph is of paramount importance in the context of our study. The operational formula for the 2-loop B-Z contributions is presented below:

\begin{equation}
\begin{split}
\Delta a_{\mu}^{(a)}&=\sum_{i}\sum_{\phi_i^0}\sum_{\mathcal{F}}\frac{\alpha_{EM} m_{\mu}^2}{4 \pi^3 v^2} N_C^{\mathcal{F}} Q_{\mathcal{F}_i}^2 \Big[y_{\phi_i^0 \mathcal{F}_i \mathcal{F}_i} y_{\phi_i \mu \mu}\mathsf{F}_{1}\Big(\frac{M_{\mathcal{F}^2}}{M_{\phi^0_i}^2}\Big)\Big]\\
	\Delta a_{\mu}^{(b)}&=\sum_{\phi_i^0}\frac{\alpha_{EM} m_{\mu}^2}{8 \pi^3 M_{\phi^0_i}^2 }  \Big[y_{\phi_i^0 \mu \mu}  \lambda_{\phi^0_i H^+H^-}\mathsf{F}_2 \Big(\frac{M_{H^{\pm}}^2}{M_{\phi^0_i}^2}\Big)\Big]\\
	\Delta a_{\mu}^{(c)}&=\sum_{\phi_i^0}\frac{\alpha_{EM} m_{\mu}^2}{8 \pi^3 v^2 }  \Big[y_{\phi_i^0 \mu \mu}  k^i_V \mathsf{F}_3 \Big(\frac{M_{H^{\pm}}^2}{M_{\phi^0_i}^2}\Big)\Big]
 \end{split}
\end{equation}

\begin{equation}
\begin{split}
\Delta a_{\mu}^{(d)}&= \frac{\alpha_{EM} m_{\mu}^2 N^{\mathcal{F}}_C}{32\pi^2 \sin\theta_{W}^2 v^2 (M_{H^{\pm}}^2 - M_W^2)} \times   \sum_{i,j=1,2}   \int_{0}^{1} dx [Q_{\mathcal{T}_i}x + Q_{\mathcal{B}_j}(1-x)] \\& \Big[y_{H^+ \mathcal{T}_i \mathcal{B}_j} \tan\beta M_{\mathcal{B}_j}^2 x (1-x) + y_{H^+ \mathcal{T}_i \mathcal{B}_j} \tan\beta M_{\mathcal{T}_j}^2 x (1+x) \Big]  \\& \times \Big[\mathsf{G}\Big(\frac{M_{\mathcal{T}_i^2}}{M_{H^{\pm}}^2},\frac{M_{\mathcal{B}_j^2}}{M_{H^{\pm}}^2}\Big) -\mathsf{G}\Big(\frac{M_{\mathcal{T}_i^2}}{M_W^2},\frac{M_{\mathcal{B}_j^2}}{M_W^2}\Big) \Big]
\end{split}
\end{equation}

These are the loop functions: 

\begin{equation}
\begin{split}
\mathsf{F}_{1}(\mathsf{\omega})& = \frac{\mathsf{\omega}}{2}\int_{0}^{1}\,dx \frac{2\mathsf{x}(1 - \mathsf{x}) - 1}{\mathsf{\omega} - \mathsf{x}(1 - \mathsf{x})}\ln\big(\frac{\mathsf{\omega}}{\mathsf{x}(1 - \mathsf{x})}\big)  \\
\mathsf{F}_{2}(\mathsf{\omega}) &= \frac{1}{2}\int_{0}^{1}\,dx \frac{\mathsf{x}(\mathsf{x} -1)}{\mathsf{\omega} - \mathsf{x}(1 - \mathsf{x})}\ln\big(\frac{\mathsf{\omega}}{\mathsf{x}(1 - \mathsf{x})}\big)
 \\
\mathsf{F}_{3}(\mathsf{\omega}) &= \frac{1}{2}\int_{0}^{1}\,dx \frac{\mathsf{x} [3\mathsf{x}(4\mathsf{x} - 1) + 10]\mathsf{\omega} - \mathsf{x}(1 - \mathsf{x})}{\mathsf{\omega} - \mathsf{x}(1 - \mathsf{x})}\ln\big(\frac{\mathsf{\omega}}{\mathsf{x}(1 - \mathsf{x})}\big) 
\end{split}
\end{equation}

\begin{equation}
\begin{split}
\mathsf{G}\big(\mathsf{\omega}^\mathsf{a},\mathsf{\omega}^\mathsf{b} \big) &= \frac{ \ln\big(\frac{\mathsf{x} \mathsf{\omega}^\mathsf{a} + \mathsf{\omega}^\mathsf{b} (1 - \mathsf{x})}{\mathsf{x}(1 - \mathsf{x})}\big)} {\mathsf{x}(1- \mathsf{x}) -  \mathsf{x}\mathsf{\omega}^\mathsf{a}  - \mathsf{\omega}^\mathsf{b} (1 - \mathsf{x})}
\end{split}
\end{equation}

\section{Results}\label{result}

This section will discuss our findings and illustrate the consequences  of the  adding VLQs to the 2HDM framework. Initially, we scrutinize the type-$X$ 2HDM, which, in the absence of VLQs, emerged as the most favorable scenario. Our analysis demonstrates a notable expansion of the pseudo scalar mass-$\tan\beta$ parameter range capable of accounting for the experimental outcome of $(g-2)_{\mu}$ .

To quantify our evaluations, we choose specific reference points for the model's parameters. These reference points adhere to all coupling limitations, align with the Higgs-to-diphoton data, and meet the constraints associated with oblique parameters, as previously elaborated in the preceding sections.

\begin{figure}[H]
	\centering
\subfloat[]	{\includegraphics[width=0.40\textwidth]{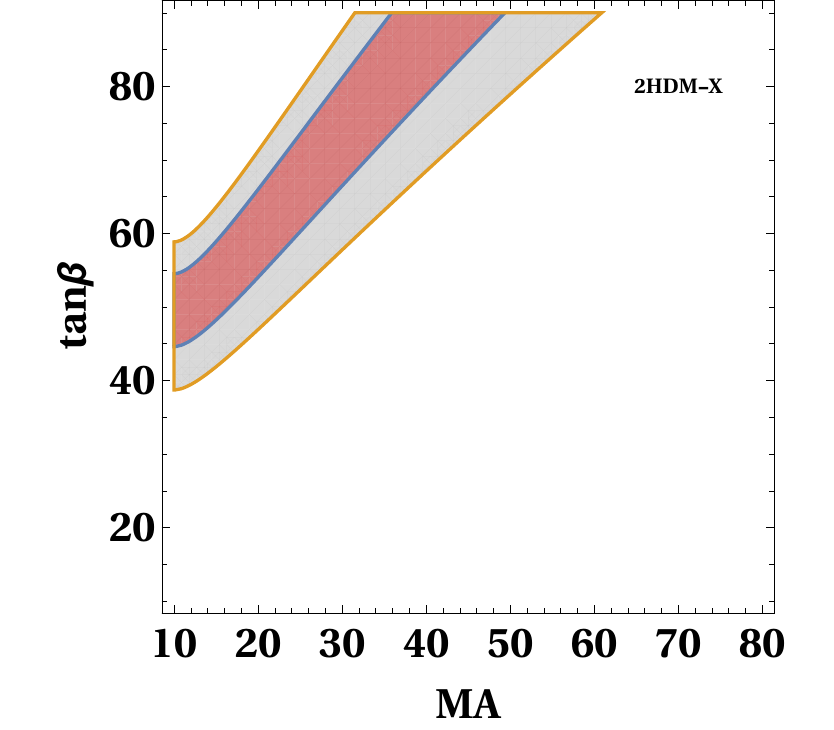}}\label{fig:A1}
\subfloat[]	{\includegraphics[width=0.40\textwidth]{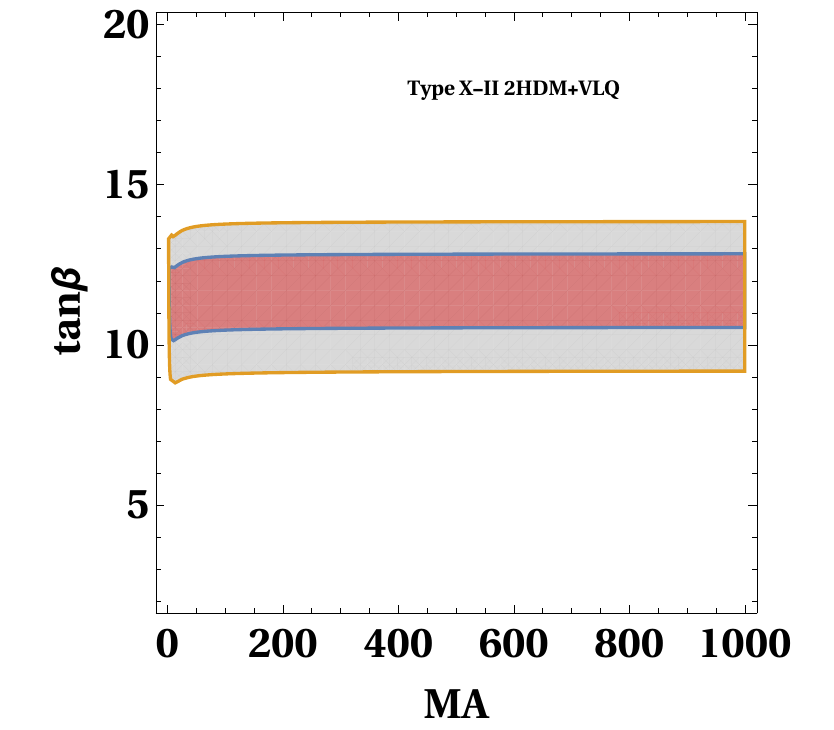}}\label{fig:B1}
	\label{fig:tanbmlme} 
\caption{{The allowed parameter space in $M_A$ - $\tan\beta$ plane for solutions $(g-2)_{\mu}$ (a) type-X 2HDM and (b) type-X-II 2HDM+VLQ, with color codes of 1$\sigma$ bright red and 2$\sigma$ (gray).}}
	\label{fig:tanbmavlq} 
\end{figure}

 First in Fig.\eqref{fig:tanbmavlq}(a) we have represented the allowed region for type-X 2HDM which is showing the highest allowed mass for $M_A$ is 60 GeV for a high value of $\tan\beta$ ($\tan\beta=80$).  The VLQs coupling with heavy Higgs in X-II 2HDM+VLQ model is  proportional to $\tan\beta$. Though the suppression factor like $\sin\theta_{\mathcal{T}}$($\sin\theta_{\mathcal{B}}$) or $\cos\theta_{\mathcal{T}}$ ($\cos\theta_{\mathcal{B}}$) is present with $\tan\beta$ but with specific choice of $\theta_\mathcal{T}$ and $\theta_\mathcal{B}$ and the mass of  VLQs the suppression effect can be alleviated. The allowed region in $M_A-\tan\beta$ plane is shown in Fig.\eqref{fig:tanbmavlq}(b) where we can see the parameter space highly expanded and we have now large values of  $M_A$ for lower $\tan\beta$ with respect to the type-X  model.

 The 1$\sigma$ and 2$\sigma$ region is shown in light red and gray color. The new VLQs play an important role in make extra contribution in $(g-2)_{\mu}$. The B-Z 2-loop contribution is shown in Fig.\eqref{fig:2loop} from (a) to (d). For the diagram (a) as the VLQs are coupled in type-II way so the down type quark coupling with heavy Higgs is $\tan\beta$ enhanced but as  $y^{\phi}_{ \mathcal{F}_1\mathcal{F}_1}=-y^{\phi}_{ \mathcal{F}_2\mathcal{F}_2}$ so these diagram contribution will be negligible. The most  contribution is  coming from the diagram (d).

The Yukawa factor $Y_{\mathcal{B}}$ is directly influenced by the mass difference between VLQs and the value of $\tan\beta$. Therefore, it is essential to carefully adjust these parameters to achieve the desired outcome. By increasing both $\tan\beta$ and the mass splitting, the Yukawa factor $Y_{\mathcal{B}}$ may exceed the perturbative limit. Thus, our approach involves selecting the largest permissible mass splitting while satisfying all other constraints, while simultaneously keeping $\tan\beta$ as minimal as feasible. Figure \eqref{fig:tanbmavlq} (b) illustrates the parameter range capable of accounting for the $(g-2)_{\mu}$ anomaly. We can substantially alter the parameter space by increasing the value of $M_A$ while consistently reducing $\tan\beta$.

\begin{figure}[H]
	\centering
\subfloat[]	{\includegraphics[width=0.40\textwidth]{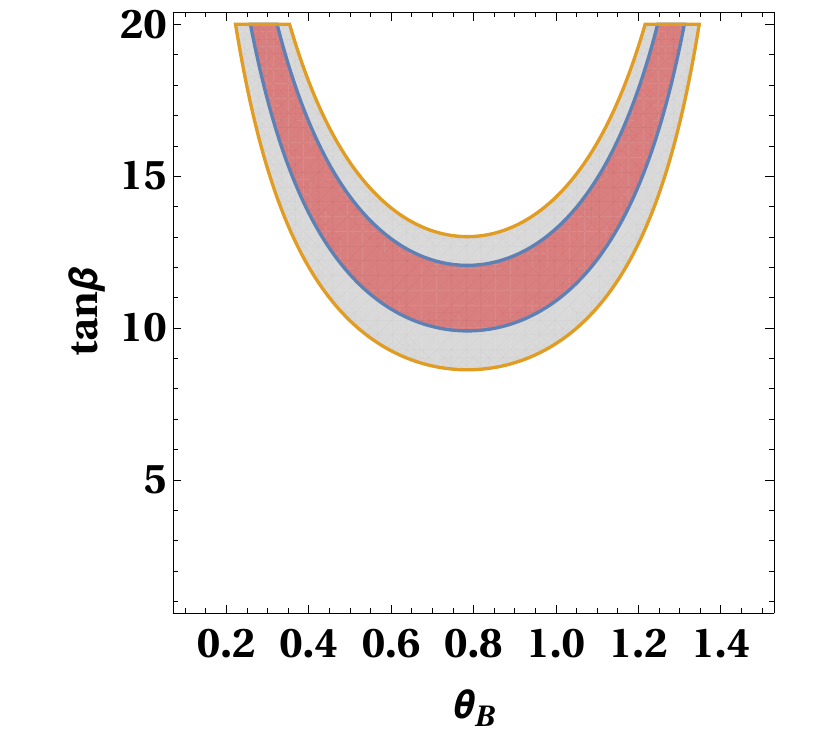}}\label{fig:A1}
\subfloat[]	{\includegraphics[width=0.40\textwidth]{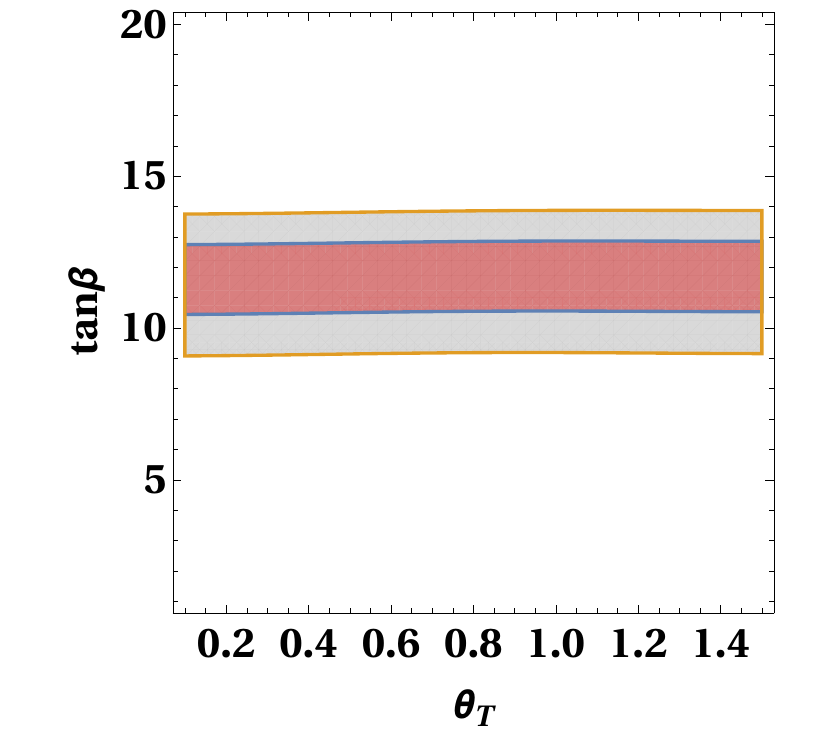}}\label{fig:B1}
	\label{fig:tanbmlme} 
\subfloat[]	{\includegraphics[width=0.40\textwidth]{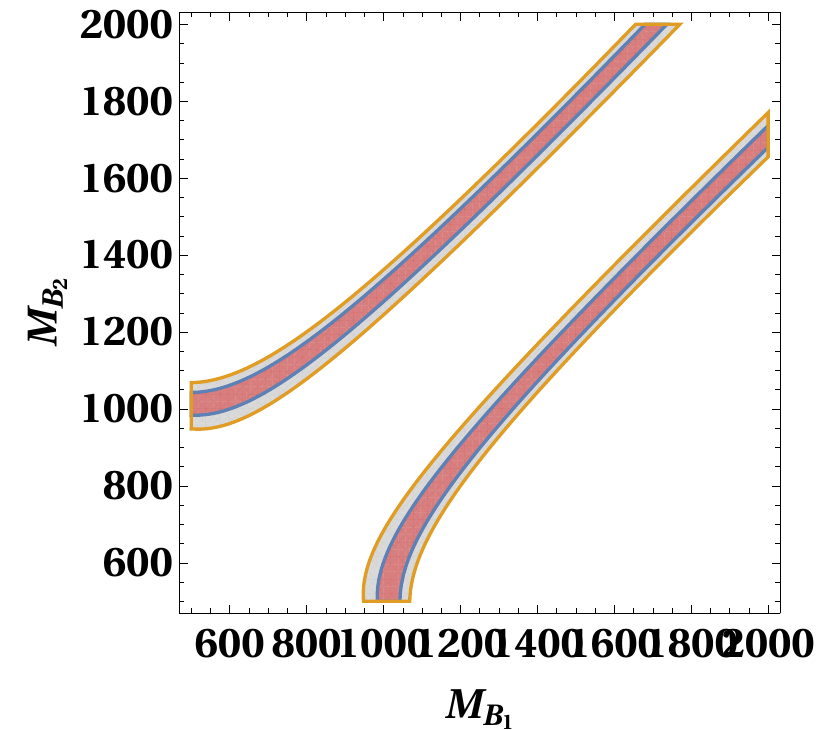}}\label{fig:A2}
\subfloat[]	{\includegraphics[width=0.40\textwidth]{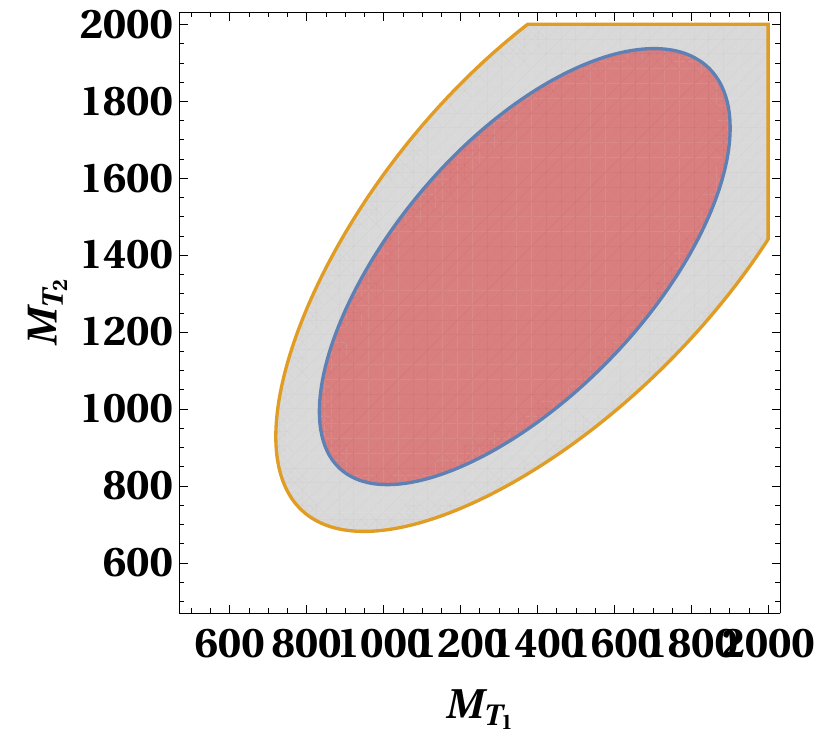}}
\label{fig:A2}
\caption{{The allowed parameter space  for solutions $(g-2)_{\mu}$ in X-II 2HDM+VLQ (a)$\theta_{\mathcal{B}}-\tan\beta $ ,(b)$\theta_{\mathcal{T}}-\tan\beta $ ,(c) $M_{\mathcal{B}_1} - M_{\mathcal{B}_2}$ and  (d)$M_{\mathcal{T}_1}-M_{\mathcal{T}_2}$ plane.}}
	\label{fig:5.6} 
\end{figure}

\begin{figure}[H]
	\centering
\subfloat[]	{\includegraphics[width=0.40\textwidth]{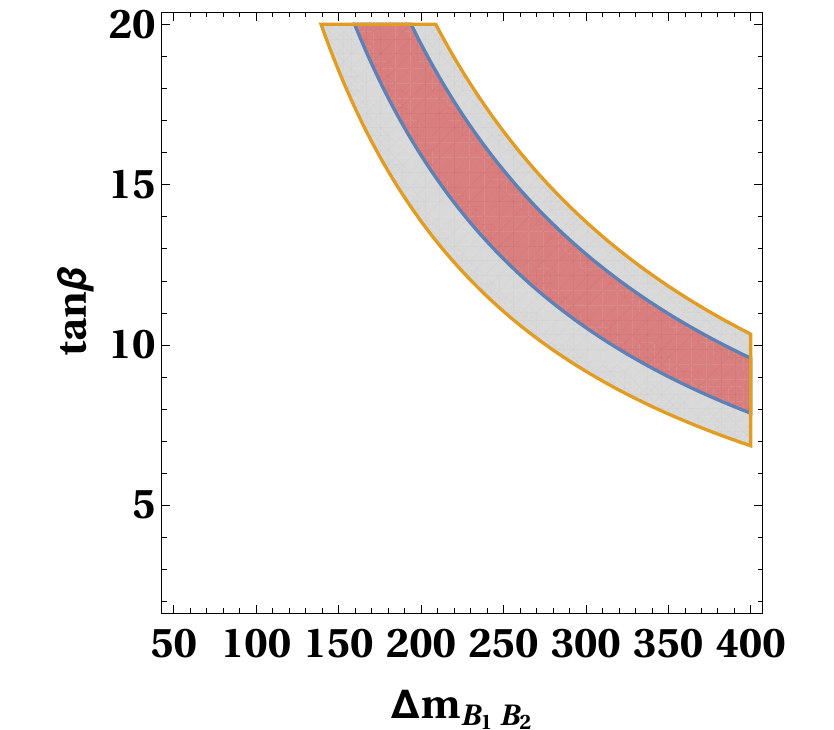}}\label{fig:A1}
\subfloat[]	{\includegraphics[width=0.40\textwidth]{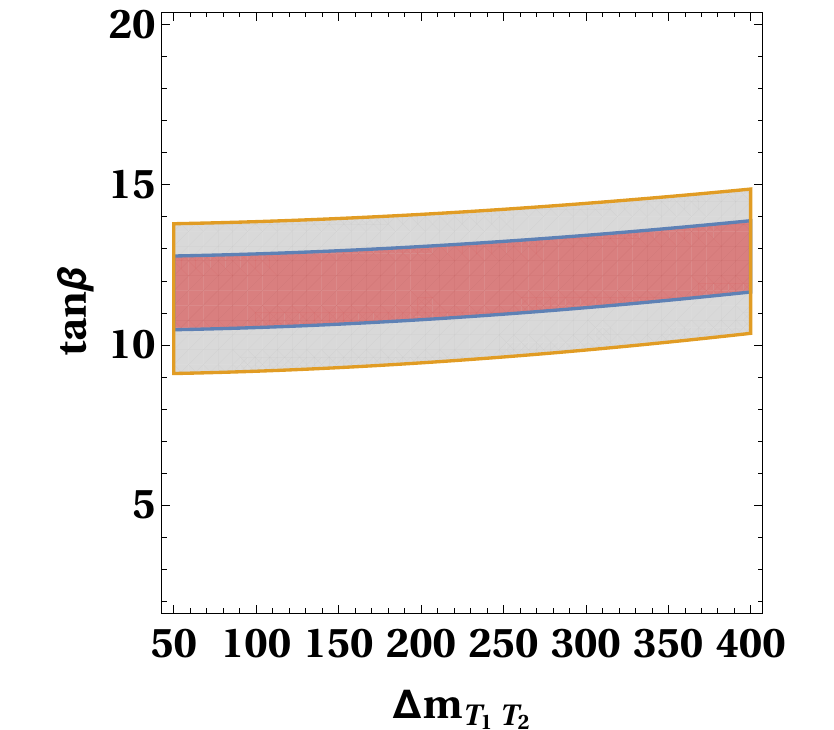}}\label{fig:B1}
\caption{{The allowed parameter space  for solutions $(g-2)_{\mu}$ in type-X-II 2HDM+VLQ (a)$\Delta m_{\mathcal{T}_1\mathcal{T}_2}-\Delta m_{\mathcal{B}_1\mathcal{B}_2}$ and(b)$\Delta m_{\mathcal{T}_1\mathcal{B}_1}-\Delta m_{\mathcal{T}_2\mathcal{B}_2}$, plane.}}
	\label{fig:5.7} 
\end{figure}

To further analyze the restrictions, we continue our investigation of the allowed parameter space satisfying $\Delta a_{\mu}$ constraints in Fig.\eqref{fig:5.6}(a-d) and in Fig.\eqref{fig:5.7}(a-b). The following graph shows our detail projection on allowed parameter space of X-II 2HDM+VLQ model.  In all these graphs we have consider the conditions (i) $M_H = M_A = M_{H^{\pm}}$,(ii) $M_{\mathcal{T}_1}>M_{\mathcal{B}_1}$, (iii) $M_{\mathcal{T}_2} <M_{\mathcal{B}_2}$ and (iv) $|Y_{\mathcal{T,B}}|<4\pi$.

 The Fig.\eqref{fig:5.6}(a-d) showing that how we can adjust the different VLQs parameters to reduce the $\tan\beta$ value for explanation of $\Delta a_{\mu}$. The mass splitting between $M_{\mathcal{B}_2}$ and $M_{\mathcal{B}_2}$ play a crucial role to reduce the $\tan\beta$ value. As the $Y_{\mathcal{B}}$
 depend on the mass splitting between $M_{\mathcal{B}_2}$ and $M_{\mathcal{B}_1}$  so we can generate a significant enhancement by increasing the mass splitting which can explain the $\Delta a_{\mu}$ as shown in Fig.\eqref{fig:5.7}(a). Our analysis have encompassed all pertinent one- and two-loop diagrams. Through this extensive investigation, we demonstrated that the parameter space allowable for the X- 2HDM is significantly expanded when considering the inclusion of VLQs, as opposed to the version of the model that excludes them.

Previously, it was established that within a X- 2HDM lacking vector-like quarks (VLQs), only very low pseudo scalar masses, specifically around $M_A < 60$ GeV, were consistent with $\Delta a_{\mu}$ for relatively high values of $\tan\beta$ (around 80). Searches for the heavier scalar where $H$ and $A$ decay into tau pairs after being created by gluon fusion are troublesome in such parameter space. Type-X 2HDM $\tau$ decays have the potential to be $\tan\beta$ improved resulting in strong signal rates that have already been ruled out by LHC collaborations \cite{ATLAS:2017eiz}. Thus, the most significant findings  is that the X-II 2HDM+VLQ model can explain the $\Delta a_{\mu}$ with expanded parameter space which can be in agreement with the heavy scalar search results at LHC.


\section{Observability of VLQ} \label{dihiggs}

\begin{figure}[H]
	\centering
\subfloat[]	{\includegraphics[width=0.48\textwidth]{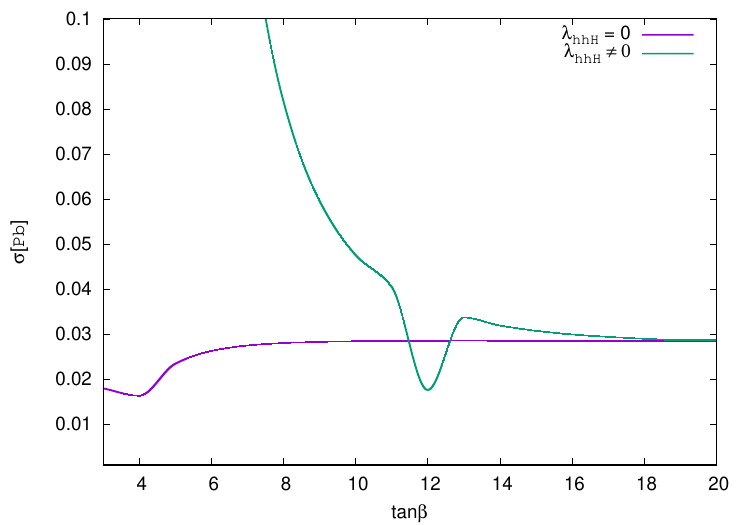} \label{dhp:a}}
\subfloat[]	{\includegraphics[width=0.48\textwidth]{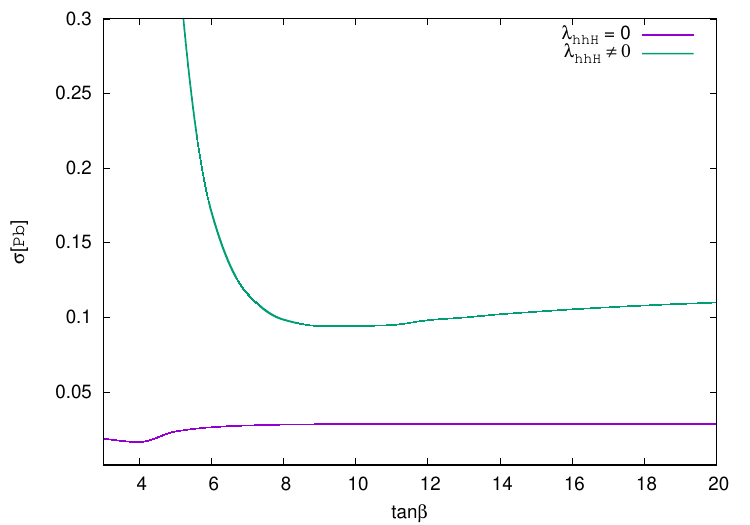} \label{dhp:b}}
\label{fig:dhp1}
\caption{(a) Double higgs production in type X-2HDM (b) Double higgs production in type X-II-2HDM+VLQs.}
	\label{fig:dhp2} 
\end{figure}

After conducting a comprehensive analysis of the anomalous magnetic moment of muon $(g-2)_{\mu}$, it prompts an inquiry into the discernible effects of the VLQs. This investigation delves into the potential influence VLQs could exert on experimental observations, adding depth to our understanding of particle physics phenomena. We will specifically look into the impact of two contributions that enter the $gg \rightarrow hh$ process of gluon fusion into Higgs pairs, which gives us direct access to $\lambda_{hhh}$ at the LHC. We will study the impact of the inclusion of a possible heavy Higgs and  VLQs contribution  with subsequent decay into a pair of the Higgs boson at 125 GeV, involving the trilinear Higgs coupling $\lambda_{hhH}$
at lowest order . We will demonstrate that the combination of the two effects has important implications on the experimental limits that can be extracted from the Higgs pair production process.

The production of DHP involves two distinct diagrams: Box and Triangle. A relatively tiny cross section of $\sigma = 31.05^{+6\%}_{-23\%}$ fb is the result of destructive interference between the triangle and box diagrams in the SM \cite{Grazzini:2018bsd,Baglio:2020wgt}. In the box diagram, only a new type of fermion can contribute, whereas in the triangular diagram, both a new fermion and a heavy Higgs (H) can participate. 

Figure \ref{dhp:a} illustrates the DHP without VLQs, utilizing only the 2HDM for two scenarios: (i) with the heavy Higgs (H) ($\lambda_{hhH} \neq 0$, shown by green solid line) and (ii) without H ($\lambda_{hhH} = 0$, shown by purple solid line). It is observed that for higher values of $\tan\beta$, both scenarios yield the DHP value predicted by the SM. Thus, solely relying on an extended scalar sector does not lead to an enhancement of the DHP.

In Figure \ref{dhp:b}, we have depicted the DHP for two scenarios, similar to the previous case, but now considering the X-II-2HDM+VLQs model. If we examine the first scenario, where the heavy Higgs (H) is absent in the DHP, it is evident from the graph that there is no increase in the DHP. However, in the second scenario where the heavy Higgs is present, the DHP cross-section is enhanced more than three times the SM value. This enhancement is solely attributed to the coupling of H with the VLQs and the non-zero value of $\lambda_{hhH}$.

Based on this comprehensive analysis, it is evident that the observation of any enhancement in the Double Higgs Production during experiments would strongly suggest the potential presence of heavy Vector-Like Quarks coupled with new scalar particles.

\section{Conclusion} \label{conclusion}
We have examined the disparity between theory and experiment concerning the anomalous magnetic moment of the muon within the framework of the 2HDM.
In the absence of vector like quarks, both the type-I and type-Y models are incapable of addressing the observed discrepancy. Furthermore, the type-II model demands the presence of light pseudoscalars, a condition that clashes with perturbative unitarity and conflicts with electroweak precision constraints. In a previous context, it was established that within a type-X 2HDM lacking vector-like quarks (VLQs), consistent alignment with $(g-2)_{\mu}$, especially at higher $\tan\beta$ values (approximately 80), was achievable only by employing notably small pseudo-scalar masses, specifically those falling within the range of $M_A < 60$ GeV. The heavier scalar, produced via gluon fusion, subsequently decays into tau pairs for both $H$ and $A$, yielding a substantial signal at the LHC. The stringent limits imposed on the parameter space by the heavy Higgs searches at the LHC play a crucial role in this context. The $\tau$ decays within the Type-X 2HDM have the capacity for enhancement with respect to $\tan\beta$, leading to robust signal rates, particularly at elevated values of $\tan\beta$. However, it's noteworthy that the parameter space supporting muon $(g – 2)_{\mu}$, which had previously been considered, has already been excluded by collaborative efforts at the LHC \cite{ATLAS:2017eiz}.Upon reexamining the limitation specific to the type-X 2HDM and taking into account the recent findings from the Muon g-2 Collaboration at Fermilab, we explored the possibility of introducing vector-like quarks into the type-X 2HDM framework. In this context, we have specifically investigated the incorporation of new vector-like quark assignments within the type-II 2HDM coupling structure. In our examination, we deliberately refrained from mixing between the vector-like quarks and the standard model quarks. 

Initially, we verified that our analysis was confined to a specific parameter space, encompassing scalar and vector-like quark (VLQ) masses, $\tan\beta$, and VLQs mixing angles. This region adheres to constraints derived from precision electroweak parameters, specifically $S$ and $T$, as well as Standard Model (SM) Higgs data. It is crucial to emphasize our commitment to ensuring compliance with Higgs decay processes, especially those involving diphotons and gluons, which are influenced by the presence of additional scalars and VLQs. Simultaneously, we maintained the coupling constants within limits that respect the principles of perturbativity, unitarity, and vacuum stability within the Higgs potential.

Our investigation encompassed a thorough analysis of $(g-2)_{\mu}$, considering all pertinent one- and two-loop diagrams. The results revealed a significant expansion of the permissible parameter space for the X-II 2HDM+VLQ compared to its counterpart lacking vector-like quarks (VLQs). In previous studies, it had been established that, within the confines of a X- 2HDM without VLQs, only very low pseudoscalar masses were compatible with $(g - 2)_{\mu}$ at high $\tan\beta$. However, our inclusion of VLQs in the analysis brought about a substantial transformation. Specifically, the pseudoscalar Higgs mass could now reach values as high as $O(1)$ TeV, even at $\tan\beta \approx 14$. 

It is well-known that double Higgs production (DHP) is the only realistic process to probe the Higgs self-coupling $\lambda$. In a CP-conserving two-Higgs-doublet model (2HDM) where all other scalars have masses of 600 GeV or more, except for the 125 GeV Higgs, DHP can investigate the couplings $\lambda_{hhh}$ and $\lambda_{hhH}$. The latter is particularly relevant in the large $\tan\beta$ limit, which enhances the coupling of the bottom VLQ to the heavier Higgs $H$. This analysis was conducted at leading order, but higher-order effects may be crucial, especially when Yukawa couplings become non-perturbative. Including such corrections is expected to increase the DHP cross-section, thereby tightening our constraints or potentially ruling out the model in extreme cases. Despite this consideration, DHP remains a vital and intriguing channel for investigation.
Before concluding, it is important to reiterate that our findings demonstrate a significant expansion of the parameter space, allowing for much larger pseudoscalar masses than previously thought possible to account for the $(g-2)_{\mu}$ anomaly. We have also demonstrated that an additional generation of vector-like quarks (VLQs) with wrong-sign Yukawa couplings, which can remain entirely undetected in single Higgs production and decay, may become observable exclusively through double Higgs production (DHP) searches.


\bibliographystyle{hephys}
\bibliography{reference}

\begin{thebibliography}{10}
\newcommand{\enquote}[1]{``#1''}

\bibitem{Muong-2:2021ojo}
B.~Abi et~al. (Muon g-2), \enquote{{Measurement of the Positive Muon Anomalous
  Magnetic Moment to 0.46 ppm}},
  \href{http://dx.doi.org/10.1103/PhysRevLett.126.141801}{\emph{Phys. Rev.
  Lett.} \textbf{126[14]} (2021) 141801},
  \href{http://arxiv.org/abs/2104.03281}{{\tt arXiv:2104.03281 [hep-ex]}}.

\bibitem{Muong-2:2021vma}
T.~Albahri et~al. (Muon g-2), \enquote{{Measurement of the anomalous precession
  frequency of the muon in the Fermilab Muon $g−2$ Experiment}},
  \href{http://dx.doi.org/10.1103/PhysRevD.103.072002}{\emph{Phys. Rev. D}
  \textbf{103[7]} (2021) 072002}, \href{http://arxiv.org/abs/2104.03247}{{\tt
  arXiv:2104.03247 [hep-ex]}}.

\bibitem{Muong-2:2021ovs}
T.~Albahri et~al. (Muon g-2), \enquote{{Magnetic-field measurement and analysis
  for the Muon $g−2$ Experiment at Fermilab}},
  \href{http://dx.doi.org/10.1103/PhysRevA.103.042208}{\emph{Phys. Rev. A}
  \textbf{103[4]} (2021) 042208}, \href{http://arxiv.org/abs/2104.03201}{{\tt
  arXiv:2104.03201 [hep-ex]}}.

\bibitem{Muong-2:2021xzz}
T.~Albahri et~al. (Muon g-2), \enquote{{Beam dynamics corrections to the Run-1
  measurement of the muon anomalous magnetic moment at Fermilab}},
  \href{http://dx.doi.org/10.1103/PhysRevAccelBeams.24.044002}{\emph{Phys. Rev.
  Accel. Beams} \textbf{24[4]} (2021) 044002},
  \href{http://arxiv.org/abs/2104.03240}{{\tt arXiv:2104.03240
  [physics.acc-ph]}}.

\bibitem{Muong-2:2006rrc}
G.~W. Bennett et~al. (Muon g-2), \enquote{{Final Report of the Muon E821
  Anomalous Magnetic Moment Measurement at BNL}},
  \href{http://dx.doi.org/10.1103/PhysRevD.73.072003}{\emph{Phys. Rev. D}
  \textbf{73} (2006) 072003}, \href{http://arxiv.org/abs/hep-ex/0602035}{{\tt
  arXiv:hep-ex/0602035}}.

\bibitem{Aoyama:2020ynm}
T.~Aoyama et~al., \enquote{{The anomalous magnetic moment of the muon in the
  Standard Model}},
  \href{http://dx.doi.org/10.1016/j.physrep.2020.07.006}{\emph{Phys. Rept.}
  \textbf{887} (2020) 1}, \href{http://arxiv.org/abs/2006.04822}{{\tt
  arXiv:2006.04822 [hep-ph]}}.

\bibitem{Joglekar:2013zya}
A.~Joglekar, P.~Schwaller and C.~E.~M. Wagner, \enquote{{A Supersymmetric
  Theory of Vector-like Leptons}},
  \href{http://dx.doi.org/10.1007/JHEP07(2013)046}{\emph{JHEP} \textbf{07}
  (2013) 046}, \href{http://arxiv.org/abs/1303.2969}{{\tt arXiv:1303.2969
  [hep-ph]}}.

\bibitem{Kyae:2013hda}
B.~Kyae and C.~S. Shin, \enquote{{Vector-like leptons and extra gauge symmetry
  for the natural Higgs boson}},
  \href{http://dx.doi.org/10.1007/JHEP06(2013)102}{\emph{JHEP} \textbf{06}
  (2013) 102}, \href{http://arxiv.org/abs/1303.6703}{{\tt arXiv:1303.6703
  [hep-ph]}}.

\bibitem{PhysRevLett.44.912}
R.~N. Mohapatra and G.~Senjanovi\ifmmode~\acute{c}\else \'{c}\fi{},
  \enquote{Neutrino Mass and Spontaneous Parity Nonconservation},
  \href{http://dx.doi.org/10.1103/PhysRevLett.44.912}{\emph{Phys. Rev. Lett.}
  \textbf{44} (1980) 912}.

\bibitem{Ma:2006km}
E.~Ma, \enquote{{Verifiable radiative seesaw mechanism of neutrino mass and
  dark matter}},
  \href{http://dx.doi.org/10.1103/PhysRevD.73.077301}{\emph{Phys. Rev. D}
  \textbf{73} (2006) 077301}, \href{http://arxiv.org/abs/hep-ph/0601225}{{\tt
  arXiv:hep-ph/0601225}}.

\bibitem{Long:1995ctv}
H.~N. Long, \enquote{{The 331 model with right handed neutrinos}},
  \href{http://dx.doi.org/10.1103/PhysRevD.53.437}{\emph{Phys. Rev. D}
  \textbf{53} (1996) 437}, \href{http://arxiv.org/abs/hep-ph/9504274}{{\tt
  arXiv:hep-ph/9504274}}.

\bibitem{Heeck:2010pg}
J.~Heeck and W.~Rodejohann, \enquote{{Gauged $L_\mu - L_\tau$ and different
  Muon Neutrino and Anti-Neutrino Oscillations: MINOS and beyond}},
  \href{http://dx.doi.org/10.1088/0954-3899/38/8/085005}{\emph{J. Phys. G}
  \textbf{38} (2011) 085005}, \href{http://arxiv.org/abs/1007.2655}{{\tt
  arXiv:1007.2655 [hep-ph]}}.

\bibitem{Schechter:1980gr}
J.~Schechter and J.~W.~F. Valle, \enquote{{Neutrino Masses in SU(2) x U(1)
  Theories}}, \href{http://dx.doi.org/10.1103/PhysRevD.22.2227}{\emph{Phys.
  Rev. D} \textbf{22} (1980) 2227}.

\bibitem{Zee:1985rj}
A.~Zee, \enquote{{Charged Scalar Field and Quantum Number Violations}},
  \href{http://dx.doi.org/10.1016/0370-2693(85)90625-2}{\emph{Phys. Lett. B}
  \textbf{161} (1985) 141}.

\bibitem{Babu:1988ki}
K.~S. Babu, \enquote{{Model of 'Calculable' Majorana Neutrino Masses}},
  \href{http://dx.doi.org/10.1016/0370-2693(88)91584-5}{\emph{Phys. Lett. B}
  \textbf{203} (1988) 132}.

\bibitem{Lindner:2016bgg}
M.~Lindner, M.~Platscher and F.~S. Queiroz, \enquote{{A Call for New Physics :
  The Muon Anomalous Magnetic Moment and Lepton Flavor Violation}},
  \href{http://dx.doi.org/10.1016/j.physrep.2017.12.001}{\emph{Phys. Rept.}
  \textbf{731} (2018) 1}, \href{http://arxiv.org/abs/1610.06587}{{\tt
  arXiv:1610.06587 [hep-ph]}}.

\bibitem{Dermisek:2013gta}
R.~Dermisek and A.~Raval, \enquote{{Explanation of the Muon g-2 Anomaly with
  Vectorlike Leptons and its Implications for Higgs Decays}},
  \href{http://dx.doi.org/10.1103/PhysRevD.88.013017}{\emph{Phys. Rev. D}
  \textbf{88} (2013) 013017}, \href{http://arxiv.org/abs/1305.3522}{{\tt
  arXiv:1305.3522 [hep-ph]}}.

\bibitem{Falkowski:2013jya}
A.~Falkowski, D.~M. Straub and A.~Vicente, \enquote{{Vector-like leptons: Higgs
  decays and collider phenomenology}},
  \href{http://dx.doi.org/10.1007/JHEP05(2014)092}{\emph{JHEP} \textbf{05}
  (2014) 092}, \href{http://arxiv.org/abs/1312.5329}{{\tt arXiv:1312.5329
  [hep-ph]}}.

\bibitem{Branco:2011iw}
G.~C. Branco, P.~M. Ferreira, L.~Lavoura, M.~N. Rebelo, M.~Sher and J.~P.
  Silva, \enquote{{Theory and phenomenology of two-Higgs-doublet models}},
  \href{http://dx.doi.org/10.1016/j.physrep.2012.02.002}{\emph{Phys. Rept.}
  \textbf{516} (2012) 1}, \href{http://arxiv.org/abs/1106.0034}{{\tt
  arXiv:1106.0034 [hep-ph]}}.

\bibitem{Bhattacharyya:2015nca}
G.~Bhattacharyya and D.~Das, \enquote{{Scalar sector of two-Higgs-doublet
  models: A minireview}},
  \href{http://dx.doi.org/10.1007/s12043-016-1252-4}{\emph{Pramana}
  \textbf{87[3]} (2016) 40}, \href{http://arxiv.org/abs/1507.06424}{{\tt
  arXiv:1507.06424 [hep-ph]}}.

\bibitem{PhysRevD.15.1958}
S.~L. Glashow and S.~Weinberg, \enquote{Natural conservation laws for neutral
  currents}, \href{http://dx.doi.org/10.1103/PhysRevD.15.1958}{\emph{Phys. Rev.
  D} \textbf{15} (1977) 1958}.

\bibitem{Wang:2018hnw}
L.~Wang, J.~M. Yang, M.~Zhang and Y.~Zhang, \enquote{{Revisiting
  lepton-specific 2HDM in light of muon $g−2$ anomaly}},
  \href{http://dx.doi.org/10.1016/j.physletb.2018.11.045}{\emph{Phys. Lett. B}
  \textbf{788} (2019) 519}, \href{http://arxiv.org/abs/1809.05857}{{\tt
  arXiv:1809.05857 [hep-ph]}}.

\bibitem{Broggio:2014mna}
A.~Broggio, E.~J. Chun, M.~Passera, K.~M. Patel and S.~K. Vempati,
  \enquote{{Limiting two-Higgs-doublet models}},
  \href{http://dx.doi.org/10.1007/JHEP11(2014)058}{\emph{JHEP} \textbf{11}
  (2014) 058}, \href{http://arxiv.org/abs/1409.3199}{{\tt arXiv:1409.3199
  [hep-ph]}}.

\bibitem{Chun:2015xfx}
E.~J. Chun, \enquote{{The muon g\ensuremath{-}2 in two-Higgs-doublet models}},
  \href{http://dx.doi.org/10.1051/epjconf/201611801006}{\emph{EPJ Web Conf.}
  \textbf{118} (2016) 01006}, \href{http://arxiv.org/abs/1511.05225}{{\tt
  arXiv:1511.05225 [hep-ph]}}.

\bibitem{Cao:2009as}
J.~Cao, P.~Wan, L.~Wu and J.~M. Yang, \enquote{{Lepton-Specific Two-Higgs
  Doublet Model: Experimental Constraints and Implication on Higgs
  Phenomenology}},
  \href{http://dx.doi.org/10.1103/PhysRevD.80.071701}{\emph{Phys. Rev. D}
  \textbf{80} (2009) 071701}, \href{http://arxiv.org/abs/0909.5148}{{\tt
  arXiv:0909.5148 [hep-ph]}}.

\bibitem{Wang:2014sda}
L.~Wang and X.-F. Han, \enquote{{A light pseudoscalar of 2HDM confronted with
  muon g-2 and experimental constraints}},
  \href{http://dx.doi.org/10.1007/JHEP05(2015)039}{\emph{JHEP} \textbf{05}
  (2015) 039}, \href{http://arxiv.org/abs/1412.4874}{{\tt arXiv:1412.4874
  [hep-ph]}}.

\bibitem{Abe:2015oca}
T.~Abe, R.~Sato and K.~Yagyu, \enquote{{Lepton-specific two Higgs doublet model
  as a solution of muon g \ensuremath{-} 2 anomaly}},
  \href{http://dx.doi.org/10.1007/JHEP07(2015)064}{\emph{JHEP} \textbf{07}
  (2015) 064}, \href{http://arxiv.org/abs/1504.07059}{{\tt arXiv:1504.07059
  [hep-ph]}}.

\bibitem{Chun:2015hsa}
E.~J. Chun, Z.~Kang, M.~Takeuchi and Y.-L.~S. Tsai, \enquote{{LHC
  \ensuremath{\tau}-rich tests of lepton-specific 2HDM for (g \ensuremath{-}
  2)$_{\mu}$}}, \href{http://dx.doi.org/10.1007/JHEP11(2015)099}{\emph{JHEP}
  \textbf{11} (2015) 099}, \href{http://arxiv.org/abs/1507.08067}{{\tt
  arXiv:1507.08067 [hep-ph]}}.

\bibitem{Gunion:2002zf}
J.~F. Gunion and H.~E. Haber, \enquote{{The CP conserving two Higgs doublet
  model: The Approach to the decoupling limit}},
  \href{http://dx.doi.org/10.1103/PhysRevD.67.075019}{\emph{Phys. Rev. D}
  \textbf{67} (2003) 075019}, \href{http://arxiv.org/abs/hep-ph/0207010}{{\tt
  arXiv:hep-ph/0207010}}.

\bibitem{Carena:2013ooa}
M.~Carena, I.~Low, N.~R. Shah and C.~E.~M. Wagner, \enquote{{Impersonating the
  Standard Model Higgs Boson: Alignment without Decoupling}},
  \href{http://dx.doi.org/10.1007/JHEP04(2014)015}{\emph{JHEP} \textbf{04}
  (2014) 015}, \href{http://arxiv.org/abs/1310.2248}{{\tt arXiv:1310.2248
  [hep-ph]}}.

\bibitem{Bhattacharyya:2014oka}
G.~Bhattacharyya and D.~Das, \enquote{{Nondecoupling of charged scalars in
  Higgs decay to two photons and symmetries of the scalar potential}},
  \href{http://dx.doi.org/10.1103/PhysRevD.91.015005}{\emph{Phys. Rev. D}
  \textbf{91} (2015) 015005}, \href{http://arxiv.org/abs/1408.6133}{{\tt
  arXiv:1408.6133 [hep-ph]}}.

\bibitem{Bhattacharyya:2013rya}
G.~Bhattacharyya, D.~Das, P.~B. Pal and M.~N. Rebelo, \enquote{{Scalar sector
  properties of two-Higgs-doublet models with a global U(1) symmetry}},
  \href{http://dx.doi.org/10.1007/JHEP10(2013)081}{\emph{JHEP} \textbf{10}
  (2013) 081}, \href{http://arxiv.org/abs/1308.4297}{{\tt arXiv:1308.4297
  [hep-ph]}}.

\bibitem{Fontes:2014tga}
D.~Fontes, J.~C. Rom\~ao and J.~a.~P. Silva, \enquote{{A reappraisal of the
  wrong-sign $hb\overline{b}$ coupling and the study of $h \rightarrow Z
  \gamma$}}, \href{http://dx.doi.org/10.1103/PhysRevD.90.015021}{\emph{Phys.
  Rev. D} \textbf{90[1]} (2014) 015021},
  \href{http://arxiv.org/abs/1406.6080}{{\tt arXiv:1406.6080 [hep-ph]}}.

\bibitem{Ferreira:2014dya}
P.~M. Ferreira, R.~Guedes, M.~O.~P. Sampaio and R.~Santos, \enquote{{Wrong sign
  and symmetric limits and non-decoupling in 2HDMs}},
  \href{http://dx.doi.org/10.1007/JHEP12(2014)067}{\emph{JHEP} \textbf{12}
  (2014) 067}, \href{http://arxiv.org/abs/1409.6723}{{\tt arXiv:1409.6723
  [hep-ph]}}.

\bibitem{Biswas:2015zgk}
A.~Biswas and A.~Lahiri, \enquote{{Alignment, reverse alignment, and wrong sign
  Yukawa couplings in two Higgs doublet models}},
  \href{http://dx.doi.org/10.1103/PhysRevD.93.115017}{\emph{Phys. Rev. D}
  \textbf{93[11]} (2016) 115017}, \href{http://arxiv.org/abs/1511.07159}{{\tt
  arXiv:1511.07159 [hep-ph]}}.

\bibitem{Ferreira:2014naa}
P.~M. Ferreira, J.~F. Gunion, H.~E. Haber and R.~Santos, \enquote{{Probing
  wrong-sign Yukawa couplings at the LHC and a future linear collider}},
  \href{http://dx.doi.org/10.1103/PhysRevD.89.115003}{\emph{Phys. Rev. D}
  \textbf{89[11]} (2014) 115003}, \href{http://arxiv.org/abs/1403.4736}{{\tt
  arXiv:1403.4736 [hep-ph]}}.

\bibitem{Das:2015mwa}
D.~Das and I.~Saha, \enquote{{Search for a stable alignment limit in
  two-Higgs-doublet models}},
  \href{http://dx.doi.org/10.1103/PhysRevD.91.095024}{\emph{Phys. Rev. D}
  \textbf{91[9]} (2015) 095024}, \href{http://arxiv.org/abs/1503.02135}{{\tt
  arXiv:1503.02135 [hep-ph]}}.

\bibitem{Ginzburg:2005dt}
I.~F. Ginzburg and I.~P. Ivanov, \enquote{{Tree-level unitarity constraints in
  the most general 2HDM}},
  \href{http://dx.doi.org/10.1103/PhysRevD.72.115010}{\emph{Phys. Rev. D}
  \textbf{72} (2005) 115010}, \href{http://arxiv.org/abs/hep-ph/0508020}{{\tt
  arXiv:hep-ph/0508020}}.

\bibitem{Horejsi:2005da}
J.~Horejsi and M.~Kladiva, \enquote{{Tree-unitarity bounds for THDM Higgs
  masses revisited}},
  \href{http://dx.doi.org/10.1140/epjc/s2006-02472-3}{\emph{Eur. Phys. J. C}
  \textbf{46} (2006) 81}, \href{http://arxiv.org/abs/hep-ph/0510154}{{\tt
  arXiv:hep-ph/0510154}}.

\bibitem{CDF:2011bsz}
T.~Aaltonen et~al. (CDF), \enquote{{Search for Production of Heavy Particles
  Decaying to Top Quarks and Invisible Particles in $p\bar{p}$ Collisions at
  $\sqrt{s}=1.96$ TeV}},
  \href{http://dx.doi.org/10.1103/PhysRevLett.106.191801}{\emph{Phys. Rev.
  Lett.} \textbf{106} (2011) 191801},
  \href{http://arxiv.org/abs/1103.2482}{{\tt arXiv:1103.2482 [hep-ex]}}.

\bibitem{CDF:2011buy}
T.~Aaltonen et~al. (CDF), \enquote{{Observation of the Baryonic Flavor-Changing
  Neutral Current Decay $\Lambda_{b} \to \Lambda \mu^{+} \mu^{-}$}},
  \href{http://dx.doi.org/10.1103/PhysRevLett.107.201802}{\emph{Phys. Rev.
  Lett.} \textbf{107} (2011) 201802},
  \href{http://arxiv.org/abs/1107.3753}{{\tt arXiv:1107.3753 [hep-ex]}}.

\bibitem{ATLAS:2011mda}
\enquote{{Search for Anomalous Missing ET in tt Events}}, .

\bibitem{CMS:2012dwa}
\enquote{{A search for the decays of a new heavy particle in multijet events
  with the razor variables at CMS in pp collisions at sqrt(s)=7 TeV}}, .

\bibitem{ATLAS:2018ziw}
M.~Aaboud et~al. (ATLAS), \enquote{{Combination of the searches for
  pair-produced vector-like partners of the third-generation quarks at
  $\sqrt{s} =$ 13 TeV with the ATLAS detector}},
  \href{http://dx.doi.org/10.1103/PhysRevLett.121.211801}{\emph{Phys. Rev.
  Lett.} \textbf{121[21]} (2018) 211801},
  \href{http://arxiv.org/abs/1808.02343}{{\tt arXiv:1808.02343 [hep-ex]}}.

\bibitem{Chala:2017xgc}
M.~Chala, \enquote{{Direct bounds on heavy toplike quarks with standard and
  exotic decays}},
  \href{http://dx.doi.org/10.1103/PhysRevD.96.015028}{\emph{Phys. Rev. D}
  \textbf{96[1]} (2017) 015028}, \href{http://arxiv.org/abs/1705.03013}{{\tt
  arXiv:1705.03013 [hep-ph]}}.

\bibitem{Dermisek:2014qca}
R.~Dermisek, J.~P. Hall, E.~Lunghi and S.~Shin, \enquote{{Limits on Vectorlike
  Leptons from Searches for Anomalous Production of Multi-Lepton Events}},
  \href{http://dx.doi.org/10.1007/JHEP12(2014)013}{\emph{JHEP} \textbf{12}
  (2014) 013}, \href{http://arxiv.org/abs/1408.3123}{{\tt arXiv:1408.3123
  [hep-ph]}}.

\bibitem{CMS:2018piu}
A.~M. Sirunyan et~al. (CMS), \enquote{{Measurements of Higgs boson properties
  in the diphoton decay channel in proton-proton collisions at $\sqrt{s} =$ 13
  TeV}}, \href{http://dx.doi.org/10.1007/JHEP11(2018)185}{\emph{JHEP}
  \textbf{11} (2018) 185}, \href{http://arxiv.org/abs/1804.02716}{{\tt
  arXiv:1804.02716 [hep-ex]}}.

\bibitem{Djouadi:1996yq}
A.~Djouadi, V.~Driesen, W.~Hollik and A.~Kraft, \enquote{{The Higgs photon - Z
  boson coupling revisited}},
  \href{http://dx.doi.org/10.1007/BF01245806}{\emph{Eur. Phys. J. C} \textbf{1}
  (1998) 163}, \href{http://arxiv.org/abs/hep-ph/9701342}{{\tt
  arXiv:hep-ph/9701342}}.

\bibitem{Grimus:2007if}
W.~Grimus, L.~Lavoura, O.~M. Ogreid and P.~Osland, \enquote{{A Precision
  constraint on multi-Higgs-doublet models}},
  \href{http://dx.doi.org/10.1088/0954-3899/35/7/075001}{\emph{J. Phys. G}
  \textbf{35} (2008) 075001}, \href{http://arxiv.org/abs/0711.4022}{{\tt
  arXiv:0711.4022 [hep-ph]}}.

\bibitem{Grimus:2008nb}
W.~Grimus, L.~Lavoura, O.~M. Ogreid and P.~Osland, \enquote{{The Oblique
  parameters in multi-Higgs-doublet models}},
  \href{http://dx.doi.org/10.1016/j.nuclphysb.2008.04.019}{\emph{Nucl. Phys. B}
  \textbf{801} (2008) 81}, \href{http://arxiv.org/abs/0802.4353}{{\tt
  arXiv:0802.4353 [hep-ph]}}.

\bibitem{Anastasiou:2009rv}
C.~Anastasiou, E.~Furlan and J.~Santiago, \enquote{{Realistic Composite Higgs
  Models}}, \href{http://dx.doi.org/10.1103/PhysRevD.79.075003}{\emph{Phys.
  Rev. D} \textbf{79} (2009) 075003},
  \href{http://arxiv.org/abs/0901.2117}{{\tt arXiv:0901.2117 [hep-ph]}}.

\bibitem{Lavoura:1992np}
L.~Lavoura and J.~P. Silva, \enquote{{The Oblique corrections from vector -
  like singlet and doublet quarks}},
  \href{http://dx.doi.org/10.1103/PhysRevD.47.2046}{\emph{Phys. Rev. D}
  \textbf{47} (1993) 2046}.

\bibitem{ATLAS:2015uhg}
G.~Aad et~al. (ATLAS), \enquote{{Search for the production of single
  vector-like and excited quarks in the $Wt$ final state in $pp$ collisions at
  $\sqrt{s}$ = 8 TeV with the ATLAS detector}},
  \href{http://dx.doi.org/10.1007/JHEP02(2016)110}{\emph{JHEP} \textbf{02}
  (2016) 110}, \href{http://arxiv.org/abs/1510.02664}{{\tt arXiv:1510.02664
  [hep-ex]}}.

\bibitem{Chen:2017hak}
C.-Y. Chen, S.~Dawson and E.~Furlan, \enquote{{Vectorlike fermions and Higgs
  effective field theory revisited}},
  \href{http://dx.doi.org/10.1103/PhysRevD.96.015006}{\emph{Phys. Rev. D}
  \textbf{96[1]} (2017) 015006}, \href{http://arxiv.org/abs/1703.06134}{{\tt
  arXiv:1703.06134 [hep-ph]}}.

\bibitem{Jegerlehner:2009ry}
F.~Jegerlehner and A.~Nyffeler, \enquote{{The Muon g-2}},
  \href{http://dx.doi.org/10.1016/j.physrep.2009.04.003}{\emph{Phys. Rept.}
  \textbf{477} (2009) 1}, \href{http://arxiv.org/abs/0902.3360}{{\tt
  arXiv:0902.3360 [hep-ph]}}.

\bibitem{Leveille:1977rc}
J.~P. Leveille, \enquote{{The Second Order Weak Correction to (G-2) of the Muon
  in Arbitrary Gauge Models}},
  \href{http://dx.doi.org/10.1016/0550-3213(78)90051-2}{\emph{Nucl. Phys. B}
  \textbf{137} (1978) 63}.

\bibitem{Lynch:2001zs}
K.~R. Lynch, \enquote{{A Note on one loop electroweak contributions to g-2: A
  Companion to BUHEP-01-16}}, \href{http://arxiv.org/abs/hep-ph/0108081}{{\tt
  arXiv:hep-ph/0108081}}.

\bibitem{Ilisie:2015tra}
V.~Ilisie, \enquote{{New Barr-Zee contributions to $\mathbf{(g-2)_\mu}$ in
  two-Higgs-doublet models}},
  \href{http://dx.doi.org/10.1007/JHEP04(2015)077}{\emph{JHEP} \textbf{04}
  (2015) 077}, \href{http://arxiv.org/abs/1502.04199}{{\tt arXiv:1502.04199
  [hep-ph]}}.

\bibitem{ATLAS:2017eiz}
M.~Aaboud et~al. (ATLAS), \enquote{{Search for additional heavy neutral Higgs
  and gauge bosons in the ditau final state produced in 36 fb$^{−1}$ of pp
  collisions at $ \sqrt{s}=13 $ TeV with the ATLAS detector}},
  \href{http://dx.doi.org/10.1007/JHEP01(2018)055}{\emph{JHEP} \textbf{01}
  (2018) 055}, \href{http://arxiv.org/abs/1709.07242}{{\tt arXiv:1709.07242
  [hep-ex]}}.

\bibitem{Grazzini:2018bsd}
M.~Grazzini, G.~Heinrich, S.~Jones, S.~Kallweit, M.~Kerner, J.~M. Lindert and
  J.~Mazzitelli, \enquote{{Higgs boson pair production at NNLO with top quark
  mass effects}}, \href{http://dx.doi.org/10.1007/JHEP05(2018)059}{\emph{JHEP}
  \textbf{05} (2018) 059}, \href{http://arxiv.org/abs/1803.02463}{{\tt
  arXiv:1803.02463 [hep-ph]}}.

\bibitem{Baglio:2020wgt}
J.~Baglio, F.~Campanario, S.~Glaus, M.~M\"uhlleitner, J.~Ronca and M.~Spira,
  \enquote{{$gg\to HH$ : Combined uncertainties}},
  \href{http://dx.doi.org/10.1103/PhysRevD.103.056002}{\emph{Phys. Rev. D}
  \textbf{103[5]} (2021) 056002}, \href{http://arxiv.org/abs/2008.11626}{{\tt
  arXiv:2008.11626 [hep-ph]}}.

\end{thebibliography}

\end{document}